\documentclass[oldversion]{aa}
\usepackage{graphicx}
\usepackage{txfonts}

\usepackage{natbib} 
\bibpunct{(}{)}{;}{a}{}{,}

\begin{document}

 \title{Magnetic properties of G-band bright points in a sunspot moat}

   \author{C. Beck\inst{1} \and L.R.\ Bellot Rubio\inst{2,1} \and
   R.Schlichenmaier\inst{1} \and P.\ S\"utterlin\inst{3}}
        
   \titlerunning{Magnetic properties of G-band bright points}
  \authorrunning{C. Beck et al.}  
\offprints{C. Beck}

   \institute{Kiepenheuer-Institut f\"ur Sonnenphysik,
              Freiburg, Germany.
	      \and Instituto de Astrof\'{\i}sica de Andaluc\'{\i}a
              (CSIC), Granada, Spain 
              \and Sterrekundig Instituut, Utrecht University, Utrecht, The Netherlands.
}
 
\date{Received xxx; accepted xxx}

\abstract{We present simultaneous spectropolarimetric observations of
four visible (630\thinspace nm) and three infrared (1565\thinspace nm)
spectral lines from the German Vacuum Tower Telescope, together with
speckle-reconstructed filtergrams in the G band and the \ion{Ca}{ii} H
line core from the Dutch Open Telescope. After alignment of the data
sets, we used the G-band intensity to locate bright points (BPs) in the
moat of a regular sunspot. With the cospatial and cotemporal
information provided by the polarimetric data, we characterize the
magnetic, kinematic, and thermal properties of the BPs. We find that
(a) 94\% of the BPs are associated with magnetic fields; (b)
their field strengths range between 500 and 1400~G, with a rather flat
distribution; (c) the contrast of BPs in the G
band depends on the angle between the vector magnetic field and the
line of sight; (d) the BPs harbor downflows of magnetized plasma and
exhibit Stokes $V$ profiles with large area and amplitude asymmetries;
(e) the magnetic interior of BPs is hotter than the immediate
field-free surroundings by about 1000~K at equal optical depth; and
(f) the mean effective diameter of BPs in our data set is 150~km, with
very few BPs larger than 300~km.  Most of these properties can be
explained by the classical magnetic flux tube model.  However, the
wide range of BP parameters found in this study indicates that not
all G-band BPs are identical to stable long-lived flux tubes or sheets
of kG strength.   }

\maketitle
\section{\label{sec_int}Introduction}
When observed at sufficient angular resolution, the solar photosphere shows
small elements that appear bright in continuum intensity.  Their
contrast is particularly high in the G band, the wavelength region
around 430~nm dominated by absorption lines of the
temperature-sensitive CH molecule, and other molecular bands such as
the CN band head at 388 nm
\citep{zakharovetal2005,uitenbroek+tritschler2006}. The discovery of
these G-band bright points (BPs) dates back to \citet{muller1983},
although they can be identified with the ``bright points'' or
``filigree'' described earlier by \citet{dunn+zirker1973} and
\citet{mehltretter1974}.

The BPs are observed preferentially along intergranular lanes, both in
active regions and in the quiet Sun. Today it is believed that they
trace magnetic flux concentrations.  The relationship between BPs and
small-scale magnetic fields has been established from observations
\citep{keller1992,berger+title2001}, magneto-hydrodynamical
simulations \citep{schuessler+etal2003,carlsson+etal2004,shelyag+etal2004}, or
semi-empirical models of flux concentrations
\citep{steiner+hauschildt+bruls2001,almeida+etal2001}. Other studies
have concentrated on the evolution of BPs
\citep{berger+title1996,vanBallegooijen+etal1998,bovelet+wiehr2003,
bonetetal2004} and on their spectral signatures
\citep{langhans+schmidt+tritschler2002,langhans+etal2004}.

Despite these advances, the magnetic field strength and other physical
properties of the G-band BPs remain largely unknown. This is because most
studies have used narrow-band filtergrams and longitudinal
magnetograms, which offer high spatial resolution but do not allow the
determination of the three components of the vector magnetic field. In this paper we investigate the magnetic, kinematic, and thermal
properties of G-band BPs in the moat of a regular spot using
simultaneous vector polarimetric measurements of four visible and
three infrared lines, complemented with diffraction-limited
narrow-band filtergrams. We also characterize the physical properties
of the outer part of the moat and surrounding quiet Sun, which do not
show G-band brightness enhancements. Our aim is to provide a complete
observational picture of the vector magnetic fields of a large
sample of BPs to help understand their nature and origin. The results
of this study may be useful for validating theoretical models and MHD
simulations of magnetic flux concentrations in the solar photosphere,
as well as to confirm the mechanisms by which G-band bright points
are generated.

The multi-wavelength observations used here have become possible only
very recently, so we give a detailed account of the reduction process
and the spatial alignment of the data sets in Sect.~\ref{observ}. Our
study is based on line parameters extracted from the observed Stokes
profiles (Sect.~\ref{dirrefpar}), as well as on the inversion of the
visible and infrared lines (Sect.~\ref{invproc}).  In Sect.\
\ref{results} we present the main results of a statistical analysis of
the field strength, field inclination, magnetic flux, velocity,
temperature, and size of BPs. Our findings are discussed in
Sect.~\ref{discussion} and summarized in Sect.~\ref{summary}.
 
\section{Observations and data reduction\label{observ}}
We observed the sunspot NOAA 10425 on August 9, 2003 from 09:36 to
10:34 UT with the two spectropolarimeters of the German Vacuum Tower
Telescope (VTT) at the Observatorio del Teide (Tenerife, Spain): the
Tenerife Infrared Polarimeter \citep[TIP;][]{martinez+etal1999} and
the Polarimetric Littrow Spectrograph
\citep[POLIS;][]{schmidt+etal2003, beck+etal2005b}. Both instruments
were used simultaneously. The same sunspot was observed between 08:25 
and 11:58 UT with the Dutch Open Telescope (DOT) at the Observatorio
del Roque de los Muchachos (La Palma, Spain). The sunspot was located
at an heliocentric angle of 27$^\circ$.

\subsection{Observational setup at the VTT}

\begin{table}
\caption{Spectral lines observed at the VTT. \label{speclines}}
\setlength{\tabcolsep}{0.6em}
\begin{tabular}{lccccr} \hline \hline
Instrument & Species & $\lambda_0$ [nm]& $\log gf$ & $g_{\rm eff}$ 
& Pixel size \cr \hline
TIP & \ion{Fe}{i} & 1564.7410  & $-0.950$ & 1.25 & 0\farcs35 \cr
TIP & \ion{Fe}{i} & 1564.8515  & $-0.669$ & 3.00 & \cr
TIP & \ion{Fe}{i} & 1565.2874  & $-0.095$ & 1.45 & \cr\hline
POLIS vis & \ion{Fe}{i} & 630.15012 &  $-0.750$ & 1.67& 0\farcs145 \cr
POLIS vis & \ion{Fe}{i} & 630.24936 &  $-1.236$ & 2.50 & \cr
POLIS vis & \ion{Fe}{i} & 630.34600 &  $-2.550$&  1.50 & \cr
POLIS vis & \ion{Ti}{i} & 630.37525 &  $-1.444$ & 0.92 & \cr\hline
POLIS UV & \ion{Ca}{ii} & 396.849 & &  & 0\farcs292 \cr \hline
\end{tabular}
\end{table}

TIP uses the main spectrograph of the VTT for observations in the infrared
(IR), while POLIS is a stand-alone spectropolarimeter with two channels. One
records the pair of \ion{Fe}{i} lines at 630 nm, and the other measures the
\ion{Ca}{ii} H line at 396 nm. In the following, the two POLIS channels will
be referred to as ``visible'' and ``UV'' channels, respectively.

An achromatic 50-50 beam splitter was used to split the light between TIP and POLIS. Scanning of the
solar surface was performed with the tip-tilt mirror of the
correlation tracker system
\citep[CT;][]{schmidt+kentischer1995,ballesteros+etal1996}, which
affects the FOV of both instruments in the same way. The CT also
provided image stabilization, allowing us to reach a spatial
resolution of about 1\arcsec, as derived from the power spectrum of a granulation area (cf.~Appendix \ref{spatres}). Both TIP and POLIS measure the polarization state by modulation of the incoming light with ferro-electric liquid crystals and a rotating waveplate, respectively. To observe the same field of view (FOV) with the two instruments, we aligned the TIP and POLIS slits by rotating
the VTT main spectrograph to the POLIS slit orientation. The lateral
displacement of the images due to the beam splitter was corrected for
with the scan mirror of POLIS.

The slit width was 0\farcs35 for TIP and 0\farcs48 for POLIS. The FOV
along the slit was 34\arcsec\,for TIP and 47\arcsec\,for the visible
channel of POLIS. The step of the spatial scan was
0\farcs35, with a total integration time of 6 seconds per scan
step. The observations presented here consist of eight scans of 70 
steps (24\farcs5) over the limb side part of the sunspot and its
surroundings. The cadence for the VTT data is about 7 minutes. 
Table \ref{speclines} summarizes the properties of the spectral lines
observed at the VTT: $\lambda_0$ is the laboratory wavelength (taken from Nave et al.\ 1994 for \ion{Fe}{i}) and $g_{\rm eff}$ the effective Land\'e factor. The pixel size along the slit is given in the last column. The full Stokes vector is available for all visible and infrared lines. For \ion{Ca}{ii} H, only the intensity profile was recorded due to the low light level. Figure
\ref{coarseali} displays maps of the observed area for the first scan
 after the coarse alignment described in Appendix \ref{dotalignment}.
\begin{figure}
\resizebox{8.8cm}{!}{\includegraphics{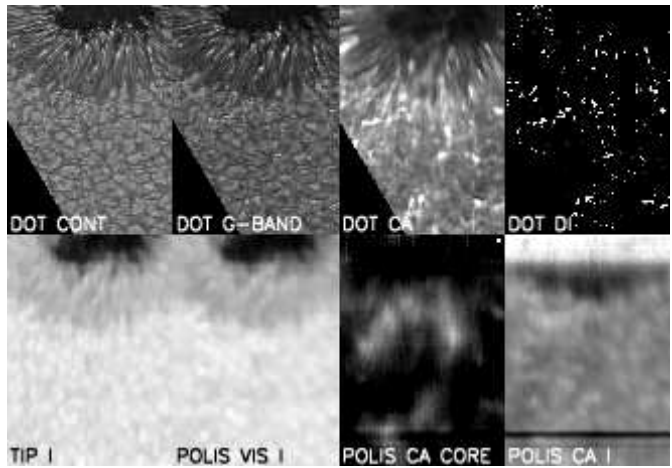}}
\caption{Maps of the first scan across the limb-side part
of NOAA 10425 after coarse alignment of DOT and VTT data. {\em Top, left
to right:} blue continuum, G-band, and \ion{Ca}{ii} H line-core filtergrams
from the DOT, and difference between the G-band and blue continuum
images. {\em Bottom, left to right:} continuum intensity at 1.5 $\mu$m, continuum intensity at 630.4\thinspace nm, intensity of the K$_{\rm 2V}$ emission peak of the \ion{Ca}{ii} H line, and intensity in the \ion{Ca}{ii} H line wing; data from the VTT.\label{coarseali}}
\end{figure}
\subsection{Observational setup at the DOT}
The data from the DOT consist of speckle-reconstructed filtergrams in
the G band at 430.5$\pm$0.5 nm, \ion{Ca}{ii} H at 396.8$\pm$0.06 nm,
and blue continuum at 432.0$\pm$0.3 nm, with a cadence of 1 minute.
The size of the FOV is 77\arcsec $\times$ 60\arcsec and the spatial
sampling 0\farcs071 per pixel. More detailed information on the image
processing and data reconstruction can be found in
\citet{suetterlin+etal2004}. The angular resolution of the filtergrams
reaches the diffraction limit of the telescope (0\farcs2 at 430.5 nm, cf.~Appendix \ref{spatres}).

\subsection{Calibration of polarimetric data}
Besides the usual dark and flatfield corrections, the polarization
measurements were corrected for instrumental polarization.  This
was done in two steps. First, TIP and POLIS have internal calibration
units in front of the CT optics. Using them, the crosstalk among the
Stokes parameters due to the optics behind the calibration units can
be determined and removed \citep{beck+etal2005b}.  The remaining
optical path consists of the telescope proper, i.e., the two coelostat
mirrors, the entrance and exit windows of the evacuated tube, the main
mirror, and one folding mirror. The instrumental polarization induced
by these elements was corrected using the telescope model of
\citet{beck+etal2005a}. Residual crosstalk between the different
Stokes parameters is estimated to be on the order of $10^{-3} I_{\rm
c}$. The rms noise in the profiles is $4 \times 10^{-4} I_{\rm
c}$ for the IR spectra and $10^{-3} I_{\rm c}$ for the visible
spectra, as evaluated in continuum windows.

\subsection{Alignment of data sets \label{alignsect_main}}
\begin{figure}
\centerline{\resizebox{8.8cm}{!}{\includegraphics{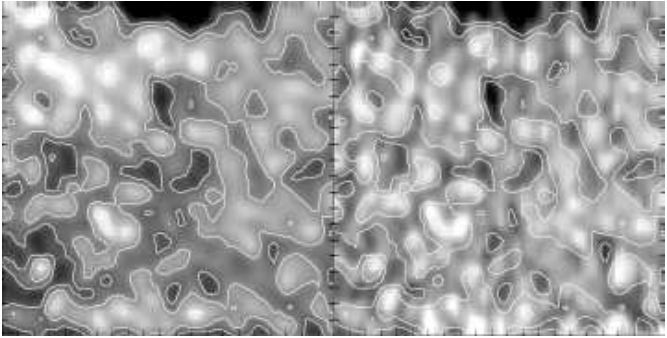}}}
\caption{Continuum intensity maps from POLIS (630 nm; {\em left}) and TIP
(1565 nm; {\em right}). Only the lower part of the FOV is shown.  Contours
outline the brightest granules and darkest intergranular lanes observed in the
POLIS map. Tick marks are separated by 1\arcsec.\label{irvisali}}
\end{figure}

For the investigation of the magnetic properties of G-band BPs,
which are identified using the DOT data, it is crucial to have a good
alignment of the data sets from the two telescopes. Additionally, the
simultaneous inversion of infrared and visible spectral lines requires
the polarization signals in both wavelength ranges to be cospatial and
cotemporal. Appendix A explains in detail the procedures we followed to align the different data sets. Figure \ref{irvisali} displays continuum
intensity maps from TIP and POLIS after coalignment. As can be seen,
the spatial correspondence of the structures in the two images is
remarkably good. The accuracy of the alignment between TIP and the
visible channel of POLIS is about 0\farcs1, or roughly a third of a
TIP pixel. The maximum displacement perpendicular to the slit due to
differential refraction \citep[see, e.g.,][]{reardon2006} was 1\arcsec, corresponding to a maximum time gap between cospatial Stokes profiles of 18 seconds (3 steps of 0\farcs35).

\begin{figure*}
\centerline{\resizebox{15.7cm}{!}{\includegraphics{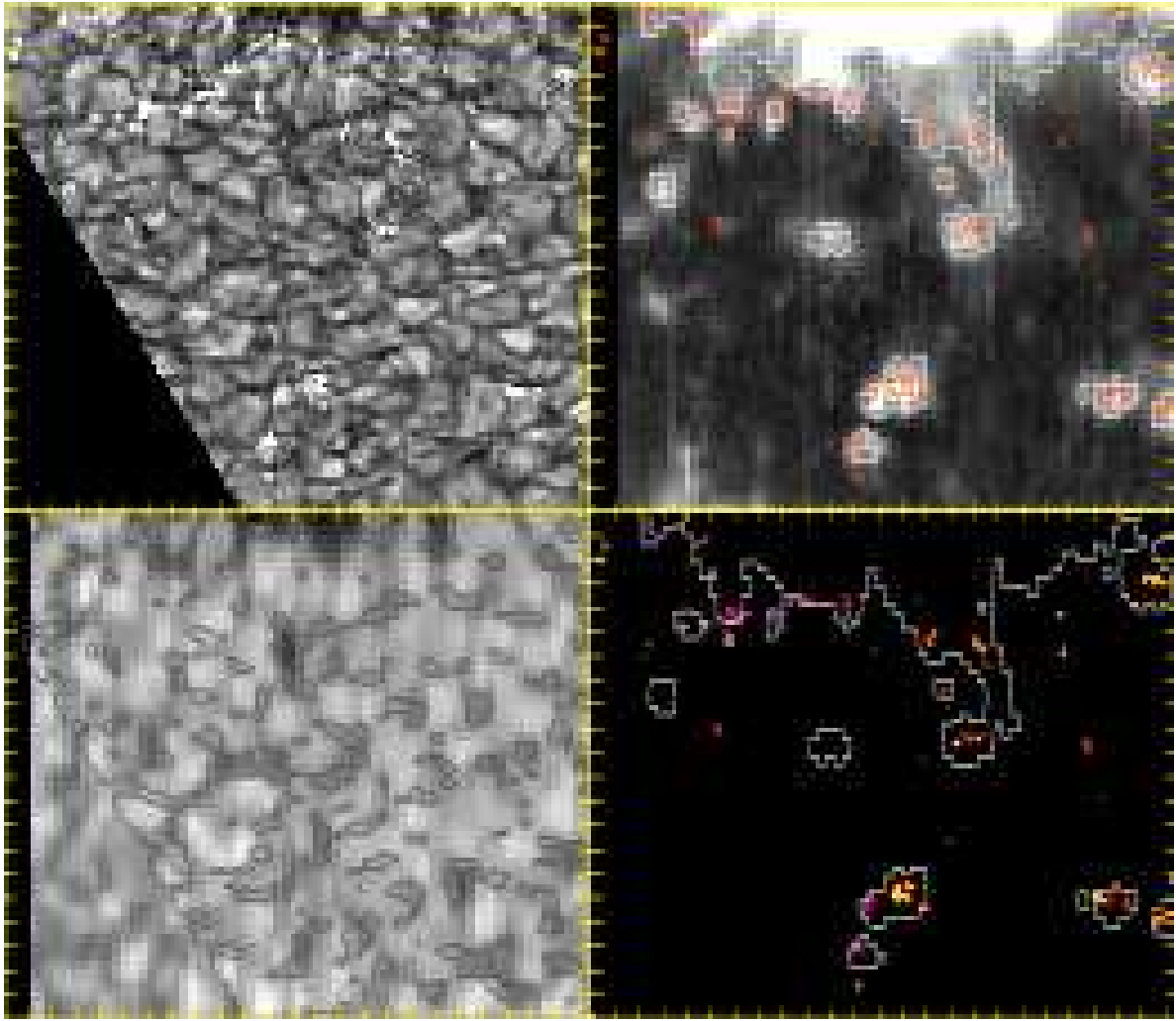}}}
\caption{{\em Left panels:} G-band intensity {\em(top)} and
IR continuum intensity {\em (bottom)} for the first
scan. The intergranular lanes visible in the G-band map can be traced
in the IR intensity map as well (black contours) despite the inferior
spatial resolution. Tick marks are separated by 1\arcsec. {\em Right
panels:} Total integrated polarization {\em (top)} and DOT BP mask
{\em (bottom)}. The individual BPs are color coded in the DOT BP
mask. White contours outline strong polarization signals ($> 1 \%$) in
the moat of the spot, as well as the sunspot canopy. Most of the
identified BPs (red contours) coincide with areas of non-zero
polarization signal ({\em top right}). \label{finalali}}
\end{figure*}

The time difference between cospatial VTT and DOT observations is 30 s or
less. The spatial misalignment between VTT and DOT data is smaller than 
one TIP pixel, i.e., 0\farcs35. This is demonstrated in
Fig.~\ref{finalali}, where we display coaligned maps for the first
scan across the FOV. Only the lower part of the FOV without the
sunspot is shown. The intergranular lanes visible in the DOT G-band
filtergram can clearly be traced in the IR continuum map. Appendix
\ref{aligneddata} contains all coaligned DOT and VTT maps.
\section{Data analysis \label{analysis}}
The cospatial and cotemporal observations described in the previous
section were used to investigate the magnetic, kinematic, and thermal
properties of G-band BPs in the moat of NOAA 10425. As a first step,
we set up a common wavelength scale for the visible and infrared
lines. Next, a number of observables were extracted from the
profiles. These line parameters, together with the results of the
inversion of the observed spectral lines, allowed us to determine the
physical properties of the various structures in the FOV. Part of our analysis is based on a statistical study that compares the properties of BPs and non-BPs (NBPs), the latter being points in the FOV that do not show enhanced G-band brightness (cf.\ Sect.~\ref{binmasks}). In this section we give details on the wavelength scale setup, the extraction of line parameters, and the inversion procedure. We also explain how BPs and NBPs were identified in the DOT filtergrams and
the VTT maps.

\subsection{Wavelength scale\label{wavelengthscale}}
The wavelength scale depends on two factors: the spectral dispersion
and the rest wavelength of the lines, $\lambda_0^i$, where index $i$
cycles through the spectral lines observed (cf.\
Table~\ref{speclines}). The dispersion affects, e.g., the
determination of the magnetic field strength from the separation of
the Stokes $V$ peaks, or any velocity defined by the relative
displacement to $\lambda_0^i$.

To set up a common wavelength scale for TIP and POLIS, we calculated a
mean quiet Sun (QS) intensity profile in the lower half of the
FOV. This was done for each of the eight scans and each spectral region
(IR or visible) separately. We then set the rest wavelengths of
\ion{Fe}{i} 1564.8~nm and \ion{Fe}{i} 630.15~nm such that the observed
positions of the line cores in the mean QS profile were at the
predicted convective blueshift values (cf.\ the last row of
Table~\ref{table_vel}). Convective blueshifts were calculated using 
the field-free, two-component QS model of
\citet{borrero+bellot2002}. Next, the dispersions were adjusted until
\ion{Fe}{i} 1565.2 nm and \ion{Fe}{i} 630.25 nm showed the correct
convective blueshifts. This procedure yielded a dispersion of 29.65
m{\AA}~pixel$^{-1}$ for the IR lines and 14.86 m{\AA}~pixel$^{-1}$ for
the visible lines. To remove the temporal evolution of the relative
motions between the Sun and the observer, a separate calibration was
used for each scan.

The measurement of Doppler shifts relative to the rest wavelength of
the lines should be accurate to $\pm 100$~m~s$^{-1}$, as this is the
uncertainty in determining line-core positions. However, the assumed
rest wavelengths (and hence the zero points of the velocity scales)
are more uncertain because of three possible influences: a reduction
of convective blueshifts due to the presence of magnetic fields, a
reduction of convective blueshifts because of the off-center position
of the spot, and a global offset due to the moat flow. The first two
effects would shift all velocities towards higher redshifts, because
we enforce the lines in the mean QS profile to have the convective
blueshift values appropriate for disc center. However, we believe the
error due to a possible reduction of convective blueshifts to be
small. First, the granulation appears to be relatively undisturbed
in the region where the average QS profiles were computed (the
moat flow prevents the formation of large-scale plage fields close 
to the spot). Second, Balthasar (1985) showed that convective
blueshifts outside disk center are reduced, but only by a small 
amount (depending on the line).

With respect to the moat flow, it is important to recall that this is
a large-scale coherent phenomenon. Since the profiles used to compute
the average QS profiles are affected by the moat flow, referring the
velocities to the mean QS profile takes out the wavelength shift 
induced by the moat flow in the same way as all other relative motions
between observer and target are removed. The only influence of the
moat flow could then be differential moat velocities across the
FOV. Assuming a horizontal moat flow of around 0.5~km~s$^{-1}$
\citep[][]{balthasar+etal1996,reza+etal2006}, we find that the maximum
error due to differential velocities cannot be larger than
$0.25$~km~s$^{-1}$ at the heliocentric angle of the observations. The
accuracy of the velocity scale thus only depends on how appropriate 
the convective blueshift values predicted by the QS model are. As 
the main effects (off-center location, presence of fields) would 
reduce the blueshift values in the QS profiles, all velocities 
derived from our wavelength scale may only be biased to show 
larger blueshifts than actually present.

We note that, prior to the determination of the wavelength scale, the
profiles from the visible channel of POLIS have been corrected for the
curvature of the spectrum along the slit. The curvature is caused by
the short focal length of the spectrograph and the large FOV along the
slit. To avoid errors in the velocity determination due to this
effect, the spectra have been shifted individually so as to have the
teluric O$_2$ 630.20 nm line (which does not show Doppler velocities)
at the same position as in the QS profile from which the
wavelength scale is determined \citep[cf.~][]{reza+etal2006}.
\begin{figure*}
\resizebox{17cm}{!}{\includegraphics{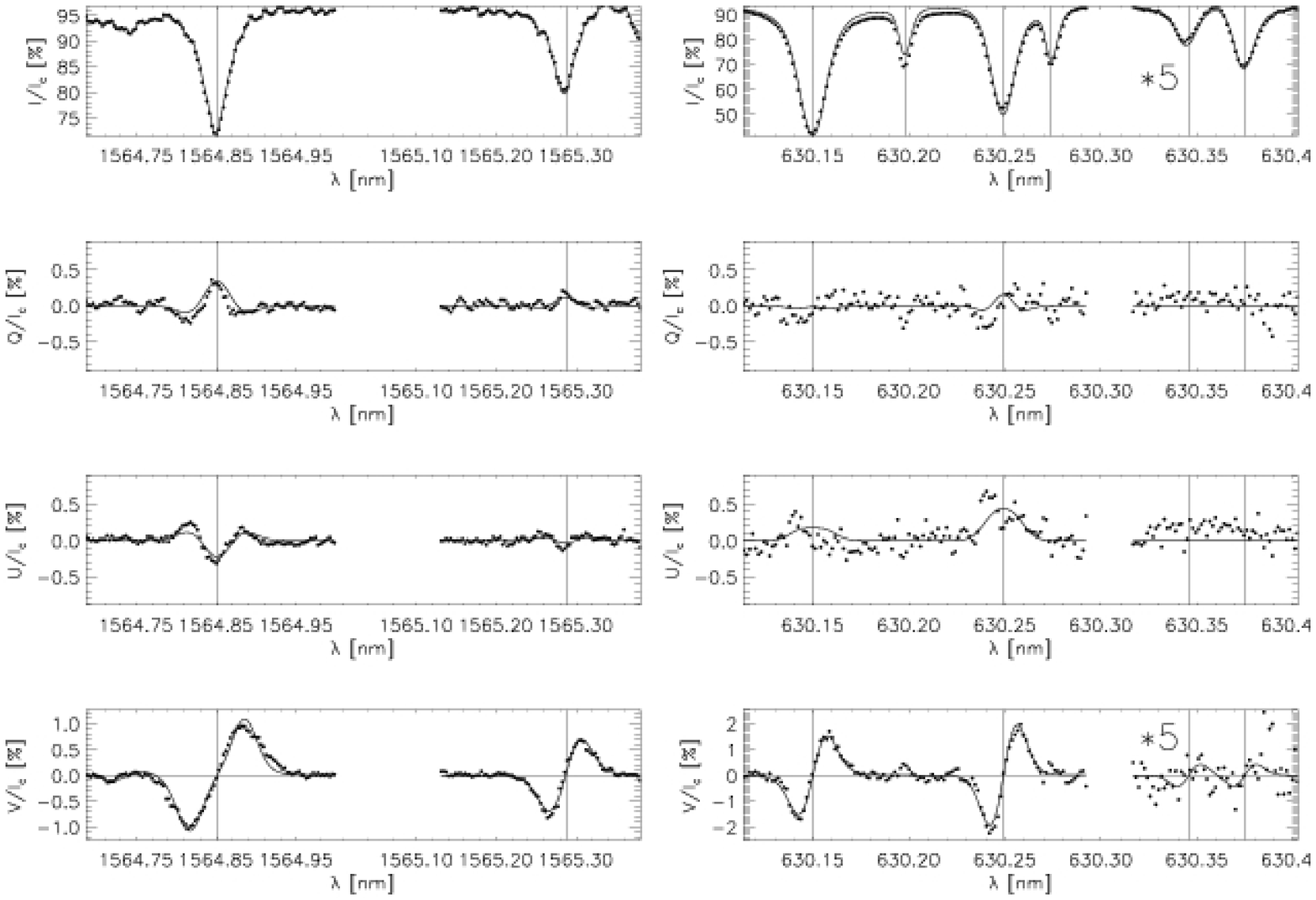}}
\caption{Example of cospatial Stokes profiles from a G-band BP,
normalized to the continuum intensity of the quiet Sun at disk
center. {\em From top to bottom:} Stokes $I$, $Q$, $U$, and $V$. The
vertical lines mark the rest wavelengths derived from the quiet Sun
intensity profiles. The observed profiles are indicated by filled
circles; the solid lines represent the best-fit profiles from the
inversion. {\em Left panel}: \ion{Fe}{i} line at 1564.8~nm with blend
at 1564.7~nm, and \ion{Fe}{i} 1565.2~nm.  {\em Right panel}:
\ion{Fe}{i} lines at 630.15 and 630.25~nm with the teluric O$_2$ lines
at 630.200 and 630.276~nm, \ion{Fe}{i} 630.35~nm and \ion{Ti}{i}
630.37~nm. The Stokes $I$ and $V$ profiles of the \ion{Ti}{i} line are
magnified by a factor of 5 for better
visibility.\label{cospatialexample}}
\end{figure*}

\subsection{Line parameters\label{dirrefpar}}
For each spatial point, we have the four Stokes profiles
of the visible and IR lines listed in Table~\ref{speclines}, the
intensity profile of the \ion{Ca}{ii} H line from the POLIS UV
channel, and the intensity in the G Band, the \ion{Ca}{ii} H line
core, and the blue continuum from the DOT, averaged over a $5 \times
5$ pixel area to simulate a TIP pixel. After alignment 
of the data sets, these observables are cospatial and cotemporal.

The four spectral lines with the highest magnetic sensitivity
(\ion{Fe}{i} 1564.8 nm and \ion{Fe}{i} 1565.2 nm in the IR,
\ion{Fe}{i} 630.15 nm and \ion{Fe}{i} 630.25 nm in the visible) 
have been used to extract the following line parameters:
\begin{itemize}
\item The line-core velocity from the intensity profiles (also 
for \ion{Ti}{i} 630.37 nm).

\item The total integrated polarization, ${\cal T}$, defined as 
\begin{equation}
\nonumber
{\cal T} = \int ({Q^2+U^2+V^2})^{1/2}/I_{\rm c} \, {\rm d}\lambda,\label{eq1}
\end{equation} 
in a wavelength range encompassing the full spectral line. 

\item The maximum polarization degree, 
\begin{equation}
\nonumber
p = {\rm max} \left\{ (Q^2+U^2+V^2)^{1/2}/I \right\}. 
\end{equation}
A profile is labeled {\em unpolarized} if the maximum polarization
degree is below 0.2\% (IR lines) or 0.5\% (visible lines), otherwise
it is called {\em polarized}. These thresholds correspond to 5 times
the noise level. Pixels with at least one spectral line above the
respective threshold were subject to a two-component inversion (cf.\
Sect.~\ref{invproc} below).
\end{itemize}
The following line parameters were calculated only for profiles 
showing polarization signal:
\begin{itemize}
\item The mean ratio of total linear and circular polarization, 
$ L/V = \langle (Q^2+U^2)^{1/2}/|V| \rangle$, in a narrow wavelength range 
around the maximum absolute Stokes $V$ signal, i.e., the ratio 
at the position of the strongest $\sigma$ component.

\item The area asymmetry of the Stokes $V$ profile, 
\begin{equation}
\nonumber
\delta A = s \times \frac{\int V(\lambda) \, {\rm d}\lambda}
{\int |V(\lambda)| \, {\rm d}\lambda}. 
\end{equation}
The polarity, $s$, is $\pm 1$ and denotes the orientation of the
magnetic field vector to the LOS. It is derived from the
position of minimum and maximum Stokes $V$ signal.

\item The number and position (with pixel accuracy) of lobes in Stokes
$Q$, $U$, and $V$. Lobes are defined as local extrema of smoothed
polarization profiles above a threshold of 0.1\% for the IR lines and
0.25\% for the visible lines.
\end{itemize}

Finally, for each Stokes $V$ profile with two distinct lobes, 
we calculated
\begin{itemize}
\item The Stokes $V$ amplitude asymmetry, 
\begin{equation}
\nonumber
\delta a = \frac{|a_{\rm blue}| - |a_{\rm red}|}{|a_{\rm blue}| + |a_{\rm red}|}, 
\end{equation}
where $a_{\rm blue}$ and $a_{\rm red}$ represent the amplitudes 
of the blue and red lobes, respectively. 
\item The wavelength position of the $V$ lobes with sub-pixel accuracy 
for the derivation of field strengths.
\item The Stokes $V$ zero-crossing velocity, $v_{\rm zcro}$, from 
a linear fit to the circular polarization signal in between the lobes. 
\end{itemize}

\subsection{Inversion procedure\label{invproc}}
The four visible and the three infrared lines observed at the VTT were
 inverted together using the SIR code \citep[Stokes Inversion
based on Response functions;][]{cobo+toroiniesta1992}.  To facilitate
the recovery of temperatures in deep photospheric layers, the Stokes
profiles of each line were normalized to the continuum intensity
of the quiet Sun at disk center. In addition, we removed the
teluric blend of the \ion{Fe}{i} 1564.8 nm line in the following
way. The \ion{Fe}{i} 1564.8 nm quiet Sun intensity profile was inverted
with a two-component quiet Sun model. The best-fit profile does not
contain the teluric blend, but reproduces the continuum outside the
spectral line. The division of the observed with the best-fit profiles
yields a multiplicative correction for the removal of the blend that
was applied to all \ion{Fe}{i} 1564.8~nm profiles.


The model atmosphere and corresponding free parameters depended on the
type of pixel inverted (Table~\ref{inv_settings}). The free parameters
in the case of field-free atmospheres were the temperature, $T$, and
line-of-sight (LOS) velocity, $v$. For magnetic atmospheres we also
had the field strength, $B$, the LOS inclination, $\gamma$, and the
azimuth of the field in the plane perpendicular to the LOS,
$\psi$. The model atmosphere was given as a function of continuum
optical depth, $\tau$, in the range from $\log \tau = 1$ to $-4$. The
temperature was determined using two nodes, i.e., the absolute value
and slope of the initial stratification could be modified. All other
atmospheric parameters were assumed to be constant with depth.  Our
choice of height-independent magnetic and kinematic parameters has
been made on purpose to facilitate the analysis of such a large data
set. Model atmospheres with height-independent parameters cannot
reproduce the asymmetries of the observed Stokes profiles, but have
been shown to provide reasonable averages of atmospheric parameters 
along the LOS.  This is exactly what is needed for a statistical 
analysis like the present one, where we focus on the general 
properties of G-band BPs rather than on the details of line formation.

The synthetic profiles were convolved with a macroturbulent velocity
$v_{\rm mac}$. In our case, $v_{\rm mac}$ describes the effects of
large-scale motions and the unknown point spread functions (PSFs) of
the spectrographs. This treatment, however, is simplistic because TIP
and POLIS have slightly different PSFs. The microturbulence, $v_{\rm
mic}$, was another free parameter of the inversion. The code determines
$v_{\rm mac}$ and $v_{\rm mic}$ separately for each inverted
pixel. However, when two-component model atmospheres were used, the
value of $v_{\rm mac}$ was forced to be the same in both
components. We employed one-component, field-free model atmospheres to invert pixels with no polarized profiles. The initial temperature stratification was
that of the Harvard Smithsonian Reference Atmosphere
\citep[HSRA,][]{gingerich+etal1971}. No stray light was allowed for. The
free parameters of the inversion were $T$, $v$, $v_{\rm mac}$, and
$v_{\rm mic}$.
\begin{table}
\caption{Free parameters for the inversion of unpolarized pixels 
{\em (second column)} and polarized pixels {\em (third and fourth
columns)}. The meaning of the symbols is as follows: x, free parameter; -, not used; 0, same value in both atmospheric
components. \label{inv_settings}}
\setlength{\tabcolsep}{0.68em}
\begin{tabular}{lccc} \hline
Parameter & Unpolarized & Polarized & Polarized \cr 
          &             & 1st comp & 2nd comp \\ \hline

Temperature ($T$)   & x & x & x \cr
Velocity ($v$)   & x & x & x \cr
Field strength ($B$)   & - & - & x \cr
Field inclination ($\gamma$)   &-  & - & x \cr
Field azimuth ($\psi$)  & - & - & x \cr
Microturbulence ($v_{\rm mic}$)   & x & x & x \cr
Macroturbulence ($v_{\rm mac}$) & x & 0 & 0 \cr
Straylight factor ($\beta$) &  - &  0 & 0 \cr
Filling factor ($f$) & - & 0 & 0 \\ \hline
\end{tabular}
\end{table}
\begin{figure}
\caption{Atmospheric parameters deduced from the inversion of the
seventh map. {\em Top:} G-band intensity and IR continuum
intensity. {\em Middle:} LOS velocity of the magnetic component and
magnetic filling factor. {\em Bottom:} Field strength and field
inclination to the LOS. White contours outline areas of enhanced
polarization signal and red contours (green in the velocity map) the
identified BPs. Tickmarks are arcsec.  Black areas represent pixels
without polarization signal that have not been inverted in terms of a
two-component model.
\label{inversionmaps}}
\end{figure}

Pixels that were classified as polarized (cf.~Sect.~\ref{dirrefpar})
were inverted with a two-component model. The first component was
identical to the field-free atmosphere described above. The second
component was magnetic. The Stokes profiles emerging from the two
components, $p_{\rm mag}$ and $p_{\rm nmag}$, are combined to yield
\begin{equation}
 p_{\rm model}(\lambda) = (1-f) \, p_{\rm nmag}(\lambda)+
f \, p_{\rm mag}(\lambda), \label{eq5}
\end{equation}
where $f$ represents the magnetic filling factor, i.e., the fractional
area of the pixel occupied by the magnetic component. Additionally,
some amount of stray light contamination was assumed. We adopted the
quiet Sun intensity profile as an approximation to the stray light
profile. In this way, the synthetic profile used to fit the
observations is
\begin{equation}
 p(\lambda) = \beta \, p_{\rm stray}(\lambda) + (1-\beta)
 \, p_{\rm model}(\lambda), \label{eq6}
\end{equation}
where $\beta$ represents the stray light factor. This combination is
intended to describe Stokes profiles emerging from unresolved flux
concentrations and their field-free surroundings, affected by 
stray light coming from unresolved granules and intergranular lanes. It is important to point out that $\beta$ and $f$ have different effects
on the position and shape of the intensity profiles, which makes it
possible to distinguish between them. The stray light profile
corresponds to an additional field-free atmospheric component with
fixed velocity and fixed temperature stratification.  In contrast, the
field-free component of the inversion can be adjusted both in velocity
and temperature to represent the physical conditions of the plasma in
the immediate vicinity of the flux concentration. We have chosen the simplest model possible for the inversion in order to have a robust estimate of magnetic properties of G-band BPs, which of course has some drawbacks. For example, the restriction to parameters that are constant in optical depth does not allow to reproduce asymmetries in the Stokes profiles.

The uncertainties in the atmospheric parameters resulting from the
inversion were estimated through the diagonal elements of the
covariance matrix \citep{press+etal1986}. Averaged over the 4000
inverted profiles of the first scan, the formal errors in $B$,
$\gamma$, $v$, and $f$ turned out to be 100~G, $10^\circ$,
400~m~s$^{-1}$, and 2\%. A number of factors may contribute to the
relatively large error in velocity: possible height-variations of the
flows that cannot be accounted for by our simple model, errors in the
convective blueshift values ($\pm 150$ m~s$^{-1}$ for each line), and
uncertainties in the laboratory wavelengths (especially for
\ion{Ti}{i} 630.37\thinspace nm). Yet, these errors are smaller than
the uncertainties associated with magnetogram or Dopplergram
observations.  The reason is the large amount of information we have
at our disposal: the four Stokes profiles of seven different lines,
which are used to constrain the model parameters simultaneously. Examples of observed and best-fit profiles are displayed in Fig.~\ref{cospatialexample}. Maps of the atmospheric parameters resulting from the inversion are shown in Fig.~\ref{inversionmaps}.
\subsection{Identification of BPs\label{binmasks}}
In order to identify BPs in the DOT data, we calculated the {\em
relative G-band enhancement} or {\em G-band contrast} map, ${\cal C}$,
using the G-band intensities, $I_{\rm GB}$, and blue continuum
intensities, $I_{\rm BC}$, as follows:
\begin{equation}
{\cal C} = \frac{I_{\rm GB}\cdot \langle I_{\rm
BC}\rangle / \langle I_{\rm GB}\rangle - I_{\rm BC}}{\langle I_{\rm
BC}\rangle} , \label{contrasteq}
\end{equation} 
where $\langle I \rangle$ denotes the spatial average of the intensity
over the map.  Only strong enhancements of the G-band intensity
relative to the blue continuum intensity are left over. An example can
be found in Fig.~\ref{coarseali}.

The BPs proper were taken to be all closed contours in the G-band
contrast map above a threshold of 0.31. This value was selected
by trial and error so as to encompass all BPs an observer would have
identified visually in the G-band images of the first scan. Our
procedure misses G-band BPs with low contrasts, but the detection of
these BPs is always problematic independent of the method used to
identify them. Contours near the upper edge of the FOV were not
considered in the analysis to avoid contamination by the sunspot
canopy. A few very short contour lines extending over 1--5 DOT pixels
were excluded manually since they are presumably due to spurious
signals created by the alignment procedure. Contours tracing G-band
brightenings inside or at the edge of granules were removed as
well. No further criterion on the contrast or the shape of the areas
was applied. The closed contours were consecutively numbered to
address individual BPs. In total, 447 separate BPs were
identified in the full data set. Our selection of bright areas by
sharp thresholding leads to extended patches, which would perhaps be
split into smaller patches by the use of more sophisticated methods like
the one proposed by \citet{bovelet+wiehr2003}.

A BP mask for the polarimetric data from the VTT was
constructed by marking all VTT pixels cospatial to one or more pixels
above the contrast threshold in the DOT data. The 447 BPs identified
in the G-band contrast maps extend over 1238 VTT pixels.  The control
sample of non-bright points (NBPs) was taken to be all VTT pixels at
least three pixels (1\farcs05) apart from any BP and outside the
canopy of the spot (visible near the upper edge of the FOV). An
example of the NPB mask is displayed in Fig.~\ref{alimapss1} for the
first map. In total, around 9500 pixels were selected for the 
NBP sample.

The majority of BPs identified by the DOT mask have sizes below
the spatial resolution of the polarimetric data. Part of the polarized
light produced by these unresolved sources will therefore be
distributed over several VTT pixels according to the spatial point
spread function. The construction of the mask for the VTT data thus
corresponds to choosing the spectra of the location where the
signature of the BP should be maximum. However, light from other
polarization sources inside the resolution element not related to a
specific BP can also contaminate the spectra attributed to the BP.
\begin{figure}
\centerline{\resizebox{7.9cm}{!}{\includegraphics{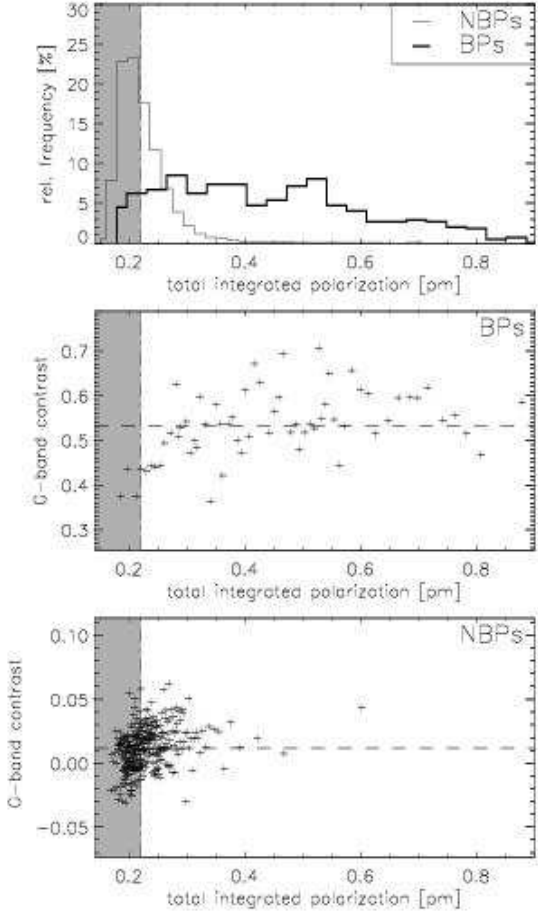}}}
\caption{ {\em Top:} Relative frequency of total integrated
polarization values for BPs {\em (thick black)} and NBPs {\em (thin
grey)}. {\em Middle and bottom:} G-band contrast vs total integrated
polarization for BPs and NBPs. The dashed horizontal lines indicate
the mean contrast of 0.52 for BPs and 0.01 for NBPs. The shaded areas
denote the range of ${\cal T}$ values within the noise
level. \label{poldeg}}
\end{figure}
\subsection{Data representation}
In the next sections, the parameters we use to describe the individual
BPs are averaged over the spatial extension of each BP in the DOT
mask and, analogously, over the corresponding area in the VTT mask if
the BP extends over more than one pixel. The NBPs always
encompass single pixels, so no averaging is carried out.

Individual G-band BPs may show different physical properties (e.g.,
magnetic flux, field strength, LOS velocity); in addition, there is
noise in the data. Both factors lead to an unavoidable scatter that
hides possible relations between the parameters characterizing the
BPs. In an attempt to reveal these relations more clearly, we have
chosen to investigate mean parameters obtained by binning the values
of individual BPs. For all scatter plots relating two quantities,
e.g., field strength and G-band contrast, we sorted the pairs
(x,y) of abscissa, x, and ordinate, y, to be in ascending order in
x. We then averaged a variable number, $N$, of values in x (typically
$N=5-50$), and averaged the values of the dependent variable y over
the same pairs. With this procedure we keep the same statistics for
each point plotted, as it is always derived from the same number of
(x,y)-pairs.  The range in x over which the averaging is performed
depends on how dense the values in x are distributed after sorting.

Thus, each data point represents an average, and the error bar
indicates the uncertainty of the mean, $\sigma/\sqrt{N}$, where
$\sigma$ is the standard deviation of the averaged values in y. The
implicit assumption here is that for a trend of, e.g., G-band contrast
with field strength, other quantities like field inclination are
randomly distributed, and their influence cancels out by the averaging
process such that only the dependence on field strength is left over. Whenever possible, the noise level is estimated from the sample of NBPs. In most cases, the corresponding results for the NPBs are  also shown for comparison. The statistics of the NPBs are better due to the larger number of data points.
\begin{table}
\caption{Fraction of unpolarized and polarized profiles of different types associated with BPs and NBPs, in percent.
\label{proftypes}} 
\setlength{\tabcolsep}{0.27em}
\begin{tabular}{lcccccc}
\multicolumn{7}{c}{BP profiles, 1238 in total} \\ \hline \hline
Line       & Unpolarized & \multicolumn{5}{c}{Polarized} \\ 
 (nm)      &        & Total & Regular & Irregular & Blue only & Red only \\ \hline
1564.8     & 3    & 97  & (35)  &   (61)  &   $<1$     &   $<1$  \\
1565.2     & 11    & 89  &  70   &    16   &   3     &   $<$1  \\
630.15     & 12   & 88  &  80   &    7   &    $<1$     &   1  \\
630.25     & 5    & 95  &  83   &    6    &   3     &   3  \\ \hline
           &        &       &         &           &           &        \\
\multicolumn{7}{c}{NBP profiles, 9497 in total} \\ \hline \hline
Line & Unpolarized & \multicolumn{5}{c}{Polarized} \\
(nm)  &             & Total & Regular & Irregular & Blue only & Red only \\ \hline
1564.8     &   33      &  67 &  (33)   &  (26)     &  6      &  2   \\
1565.2     &   72      &  28 &  15   &  4     &  7     &  2 \\
630.15     &   74      &  26 &  20   &  $<1$      &  3     &  2 \\
630.25     &   49      &  51 &  22   &  2      &  20     & 7 \\
\hline
\end{tabular}
\end{table}
\section{Results\label{results}}
Our results derive either directly from the spectral line profiles or
from the more complex inversion procedure. In some cases, we used both
approaches to confirm specific findings, while in other cases they were
complementary.
\subsection{Polarization signature of BPs}
A visual inspection of Figs.~\ref{finalali}, \ref{inversionmaps}, or
\ref{alimapss1} suggests that the majority of G-band BPs are
associated with polarization signals. To quantify this impression, we
use the total integrated polarization, ${\cal T}$.  ${\cal T}$ is
calculated separately for the four most Zeeman-sensitive spectral
lines according to Eq.~(\ref{eq1}), and then averaged over all
lines. We estimate the noise level in ${\cal T}$ by averaging ${\cal T}$ over
all profiles that were not inverted due to their maximum polarization
degrees being below the corresponding threshold. The average value
plus one standard deviation, ${\cal T}_{\rm mean} + \sigma$, is
0.22 pm for these profiles. Values below this limit are assumed
to correspond to the integration of pure noise, where of course we
cannot exclude the existence of weak polarization signals below the
noise level. The histogram of ${\cal T}$ values (upper panel of
Fig.~\ref{poldeg}) demonstrates that the majority of BPs (94\%) have
polarization signals above this level, whereas half of the NBPs have
${\cal T}$ values below it. Thus, clear polarization signals (i.e.,
magnetic fields) are observed no farther than 0\farcs35 from the
position of almost all G-band brightenings. Due to the 1\arcsec 
spatial resolution of the spectra we cannot exclude that the 
polarization signal comes from a source outside the pixel itself, 
but we think it highly improbable that it happens so often by chance. The middle and lower panels of Fig.~\ref{poldeg} display the G-band
contrast of BPs and NBPs as a function of ${\cal T}$. We find that the
contrast of BPs only depends on ${\cal T}$ for small polarization
signals on the order of the integrated noise. For ${\cal T}$ values
higher than 0.3 pm, the BP G-band contrast is approximately constant. The G-band contrast stays close to zero for the NBPs and shows no systematic trend with ${\cal T}$. This suggests that the NBP sample is distinctly different from the BP sample, even if it contains locations that exhibit some polarization signal.
\begin{figure}
\centerline{\resizebox{8cm}{!}{\includegraphics[bb=58 360 256
522,clip]{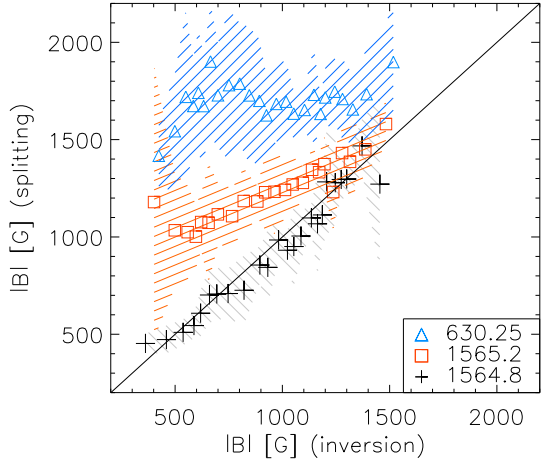}}}
\caption{Magnetic field strength of BPs from the Zeeman splitting of
\ion{Fe}{i} 1564.8 nm ($+$), 1565.2 nm ($\Box$), and 630.25 nm
($\bigtriangleup$), vs the values inferred from the inversion. The shaded
areas give the scatter inside the bins. The straight line indicates a
one-to-one correspondence. \label{scatterfldstrg}}
\end{figure}

The shape of the Stokes $V$ profiles gives
information about the geometry of the magnetic field and the dynamical
state of the flux concentrations. After the distinction between polarized and unpolarized profiles, the polarized profiles (cf.~Sect.~\ref{dirrefpar}) were classified according to the number of Stokes $V$ lobes. A profile is called regular if it has 2 lobes, irregular if it shows 3 or more lobes, and blue/red only if it shows only one blue/red lobe at shorter/longer wavelength than the intensity minimum. The \ion{Fe}{i} line at 1564.8~nm is blended with \ion{Fe}{i} 1564.7~nm, which explains the high number of irregular profiles; the respective numbers are less reliable and thus are given in parentheses (Table \ref{proftypes}). We find that roughly 80\% of the BP profiles are regular with two lobes in Stokes~$V$, while 10 \% are irregular with either one or three lobes, or unpolarized (10\%). In the sample of NBPs, the fraction of unpolarized and irregular Stokes~$V$ profiles is significantly larger. A high percentage of BPs are associated with regular two-lobed Stokes $V$
profiles. There are few cases of irregular three-lobed profiles, which
would indicate fields with opposite polarity and different LOS
velocities in the same pixel. Examples of this kind can be found in
the sunspot penumbra at the upper boundary of the FOV. Thus, even if
the asymmetries of the Stokes $V$ profiles (cf.~Sect.~\ref{assym})
indicate some variation in the magnetic properties along the LOS, most
BPs do not seem to be related to a complex multi-component field
topology.

\begin{figure}
\centerline{\resizebox{8.2cm}{!}{\includegraphics{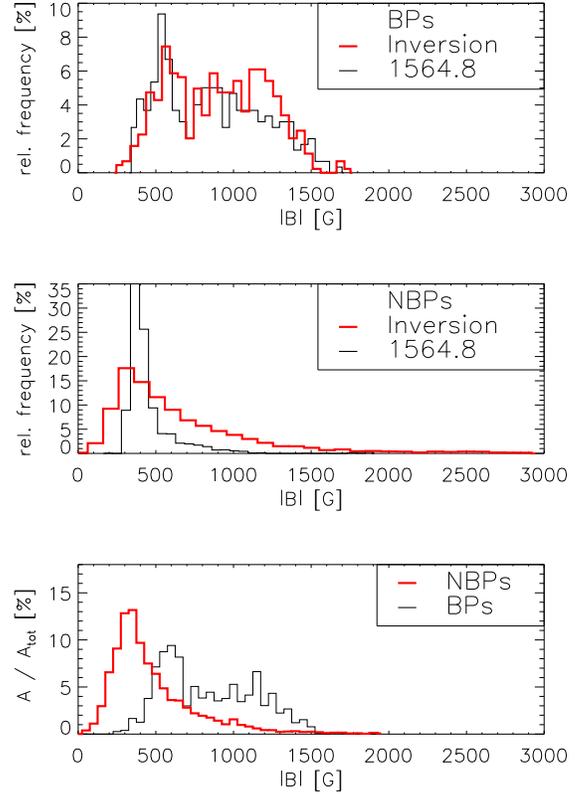}}}
\caption{{\em Top:} Distribution of BP field strengths as inferred
from the inversion (thick line) and the splitting of the \ion{Fe}{i}
1564.8 nm line (thin line). {\em Middle:} same for the NBP
sample. {\em Bottom:} Fraction of the total area occupied by
fields of given strength. \label{histfldstrg}}
\end{figure}
\subsection{Magnetic field strength \label{fieldstrength}}
For spectral lines in the strong field limit, where the separation of
the $\sigma$ components is larger than the thermal Doppler broadening,
the magnetic field strength ($B$) can be calculated from the position
of the Stokes $V$ lobes according to
\begin{equation}
\label{eq10}
\Delta \lambda = 4.67 \times 10^{-13} g_{\rm eff} B \lambda_0^2,
\label{beq}
\end {equation}
where $\Delta \lambda$ is the wavelength distance between the peaks of
Stokes $V$ and the line center. In observations of magnetic fields
outside sunspots or pores, Eq.\ (\ref{eq10}) only fully applies to the
IR line at 1564.8 nm for fields above $\sim$500~G.  However, for a
comparison with the inversion results, we calculate $B$ from
Eq.~(\ref{beq}) for the three most Zeeman-sensitive lines.

Figure \ref{scatterfldstrg} displays the BP field strengths indicated by the
splitting of the different lines versus the field strengths inferred from the
inversion. Taking the latter as the most accurate ones, it is clear that the
splitting of \ion{Fe}{i} 1564.8 nm ($+$) serves as a good diagnostics of $B$ 
for fields down to 500 G. The \ion{Fe}{i} 1565.2 nm line ($\Box$), and 
 the visible line at 630.25 nm ($\bigtriangleup$) especially, show strong 
deviations for fields below 1200 G. The same result was obtained for
the NBP sample.

\begin{figure}
\centerline{\resizebox{8.8cm}{!}{\includegraphics[bb=58 365 291
473,clip]{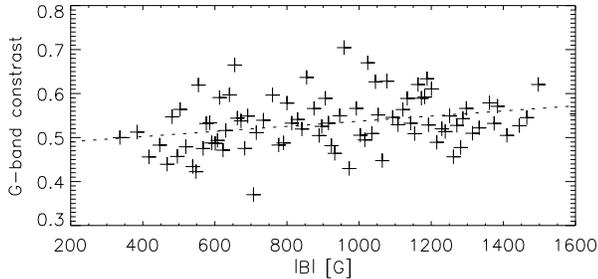}}}
\caption{G-band contrast vs field strength for the BP sample. The
dashed line represents a linear fit to the data points.\label{bvsdi}}
\end{figure}

The distribution of field strengths is displayed in
Fig.~\ref{histfldstrg} for individual BPs and NBPs. The
inversion and the splitting of \ion{Fe}{i} 1564.8 nm essentially give
the same result: the field strengths of BPs are distributed uniformly
in the range from 500~G to about 1300-1400~G. The finding of G-band
BPs with strengths of $\sim 500$~G may pose a serious problem for our
understanding of why these flux concentrations are bright. In general,
models predict zero or even negative G-band contrast for 500~G
\citep{almeida+etal2001,shelyag+etal2004}, although high positive 
contrasts are also detected.
The distribution of field strengths in the NBP control sample hints at
a quickly decreasing probability for stronger fields. The same result
has been found by \citet{khomenko+etal2003} in their study of quiet
Sun internetwork fields using IR lines. The sharp drop of the relative
frequency of fields weaker than 300 G may be imposed by the detection
limit of the observations
\citep[cf.][]{khomenko+etal2003}. Remarkably, it seems that the spot
does not influence the distribution of weak fields in the outer
part of the moat and surrounding quiet Sun much (the regions where the NBP
sample comes from, cf.~Fig.~\ref{alimapss1}). To allow comparisons with magnetoconvection simulations, the
lower panel of Fig.~\ref{histfldstrg} shows the area occupied by magnetic fields of a given strength using 100 G bins, as a fraction of the total area occupied by fields in the BP and NBP sample, respectively. Figure \ref{bvsdi} demonstrates that there is a slight increase in the G-band contrast toward higher field strengths for the BPs.
\begin{figure}[b]
\centerline{\resizebox{8.6cm}{!}{\includegraphics{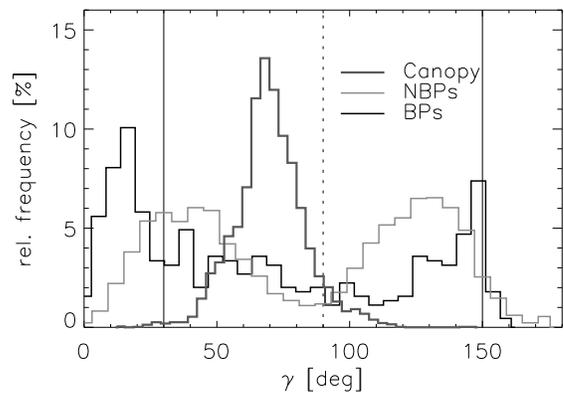}}}
\caption{Histogram of the magnetic field inclination with respect to
the LOS for BPs (black), NBPs (grey), and the canopy area close to the
sunspot (light grey). The vertical dashed line corresponds to fields
perpendicular to the LOS. The vertical solid lines mark the LOS
inclinations of fields, which are purely vertical relative to the solar
surface.\label{histinc}}
\end{figure}
\begin{figure*}
\centerline{\includegraphics[bb=60 400 525 772,clip]{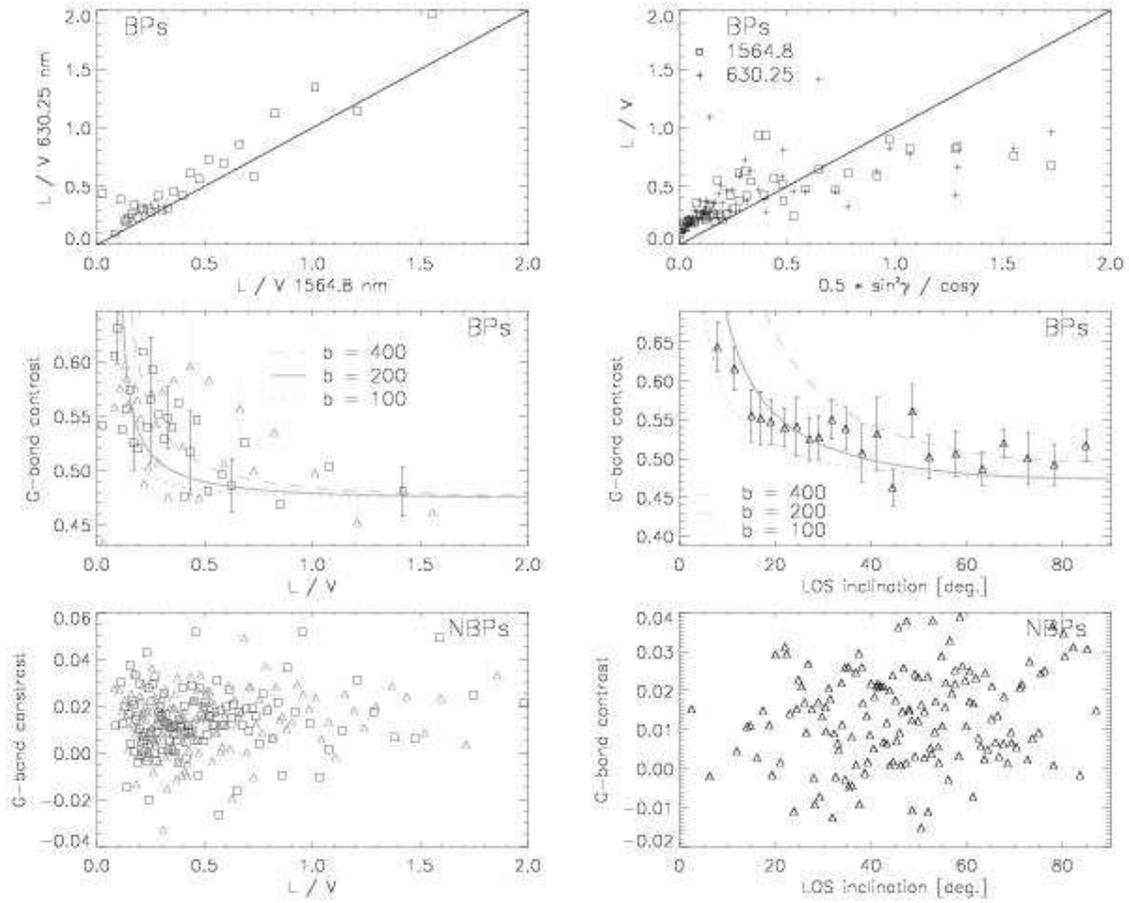}}
\caption{{\em Top panels:} $L/V$ of \ion{Fe}{i} 630.25 nm vs the $L/V$
of \ion{Fe}{i} 1564.8 nm {\em (left)}, and $L/V$ of 1564.8 nm
($\Box$) and 630.25 nm ($+$) vs the values obtained from
Eq.~(\ref{eqlpv}) by inserting the inclinations deduced from the
inversion {\em (right}).  {\em Middle left panel:} Dependence of the
BP contrast on the $L/V$ ratio of 1564.8 nm ($\bigtriangleup$) and on the $L/V$ ratio averaged over the two visible lines ($\Box$). {\em Middle right
panel:} Dependence of the BP contrast on the field inclination deduced
from the inversion. Overplotted are curves calculated from
Eq.~(\ref{empirical}) for three values of the parameter $b$. Error bars
indicate the uncertainty of the mean values.  {\em Bottom panels}:
BP contrast vs L/V and vs inclination, for NBPs.\label{lpv}}
\end{figure*}

\subsection{Magnetic field inclinations \label{inclination1}}
Figure \ref{histinc} displays histograms of the LOS inclination for
BPs and NBPs. Positive and negative polarities are found in both types
of structures. In the case of the NBPs, the inclinations are
distributed rather symmetrically around two broad peaks at $\gamma =
40^\circ$ and 130$^\circ$, with few fields perpendicular to the
line of sight. For comparison, we also show the histogram of LOS inclinations found at the upper border of the FOV, close to the penumbra and its
filaments (cf.~Fig.~\ref{alimapss1}). The fields in this area exhibit
a single peak at around 70$^\circ$, which corresponds to the unipolar
horizontal fields of the sunspot canopy. The distribution of LOS inclinations for the BPs peaks at around 15$^\circ$ and 145$^\circ$; the peaks are less broad than those of the NBP sample. Few BPs show fields perpendicular to the LOS. The LOS inclinations of BPs and NBPs at a heliocentric angle of $\theta = 27^\circ$ are consistent  to the first order with fields that are close to vertical to the surface. The systematic displacements
toward smaller inclination (BP peak at 15$^\circ$, NBP at 130$^\circ$)
could be due either to the influence of the canopy fields of the
sunspot or to unresolved mixed polarity fields \citep{lites2002}.

The inclination of the vector magnetic field to the LOS can also be
estimated from the ratio of linear to circular polarization, $L/V$. The $L/V$
ratio is more affected by noise in the data than the inclination derived 
from the inversion, especially for small amplitudes of $L$ or $V$. For 
fully split lines, \citet{landi2003} demonstrate that
\begin{equation}
L / V \simeq \frac{1}{2} \, \frac{\sin^2 \gamma}{\cos \gamma}. \label{eqlpv}
\end{equation}

The upper left panel of Fig.~\ref{lpv} displays a scatter plot of the
$L/V$ ratio for the two most Zeeman-sensitive spectral lines,
\ion{Fe}{i} 1564.8~nm and \ion{Fe}{i} 630.25~nm. The values of $L/V$
are similar in both wavelength regions; they only show significant
deviations  for values of $L/V$ above 0.45 or, according to
Eq.~(\ref{eqlpv}), for inclinations larger than 50$^\circ$. The upper
right panel of Fig.~\ref{lpv} compares the $L/V$ values with the
results of the inversion. This plot demonstrates that
Eq.~(\ref{eqlpv}) is also valid for the visible \ion{Fe}{i} 630.25~nm
line if the field is not too much inclined ($\gamma\ll 50^\circ$). In
general, the observed values of $L/V$ are slightly higher than those
indicated by the inversion, which we attribute to the influence of
noise.

Interestingly, the G-band contrast of BPs seems to depend on the
magnetic field inclination. The middle panels of Fig.~\ref{lpv} show
the contrast as a function of both $L/V$ and the magnetic field
inclination inferred from the inversion. We find a correlation of the
brightest structures with magnetic fields parallel to the LOS. The BPs
with magnetic fields perpendicular to the LOS systematically exhibit
reduced G-band contrasts. We believe this trend to be significant
even if the contrast shows some scatter due to intrinsic differences
in the properties of the BPs like field strength or flux, which also
have some effect on the contrast.  The overplotted curves will be
discussed in Sect.~\ref{discussion}. No relation between the G-band
contrast and the field inclination is observed for the NBPs 
(bottom panels of Fig.~\ref{lpv}).
\subsection{LOS velocities}
We used two proxies to estimate flows of magnetized and field-free
plasma: the Stokes $V$ zero-crossing shift and the Stokes $I$ line-core
velocity, respectively. The inversion also yielded values for the 
LOS velocity in the magnetic and field-free components of the atmosphere. 
By convention, positive velocities indicate redshifts.

\begin{figure}
\centerline{\resizebox{8cm}{!}{\includegraphics{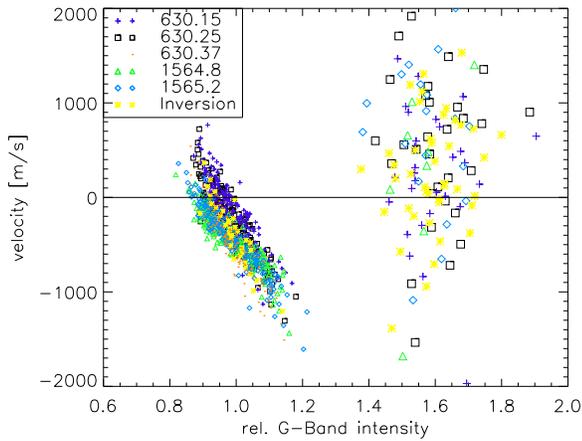}}} 
\caption{Relation between the LOS velocity and the normalized G-band
intensity. Positive velocities indicate redshifts. {\em Small
symbols:} Stokes $I$ line-core velocity and LOS velocity of the
field-free component of the inversion for the NBP sample.  {\em Large
symbols:} Stokes $V$ zero-crossing velocity and LOS velocity of the
magnetic component of the inversion for BPs.\label{intvel}}
\end{figure}

Figure \ref{intvel} shows the dependence of the LOS velocity on the
G-band intensity. For easier comparison with other data sets, we used
the relative G-band intensity $1 + {\cal C}$ in this plot. For the NBP
sample (small symbols in the left half of the figure, with relative
intensities below 1.31), we find that the velocities derived from the
Stokes $I$ line-core position and the velocities of the field-free
component of the inversion are in good agreement with the expectation
for an area dominated by granulation: bright granules show blueshifts
($v < 0$), whereas intergranular lanes have reduced intensity and are
associated with redshifts ($v> 0$).
In contrast, no trend of the magnetic velocity with intensity is
found in the BPs (large symbols, relative intensities above 1.31 in
Fig.~\ref{intvel}). The velocity of the BPs spans the range $\pm
2$~km~s$^{-1}$ for the binned data. The extreme velocities,
considering only points with a clear polarization signal above 1\%,
reach $\pm 4$~km~s$^{-1}$. In general, all spectral lines and the
inversion show a preference for downflows of magnetized plasma inside
the BPs. 

\begin{table}[b]
\caption{LOS velocities of BPs and NPBs in m~s$^{-1}$.\label{table_vel}}
\begin{tabular}{lc@{\hspace{.2cm}}c@{\hspace{.2cm}}c@{\hspace{.2cm}}c@{\hspace{.2cm}}c} 
\hline \hline
                     & 630.15 & 630.25 & 630.37 & 1564.8 & 1565.2 \cr\hline 
BPs, zero-crossing   &  300   & 540    &        & 394    &   598   \cr 
BPs, line core       & $-24$    &  $-49$    & $-200$ & $-296$ &  $-267$ \cr
NBPs, line core      & $-172$ & $-250$ & $-451$ & $-448$ &  $-447$ \cr 
Convective blueshift & $-185$ & $-262$ & $-424$ & $-445$ &  $-469$ \cr \hline
\end{tabular}
\end{table}

\begin{figure}
\centerline{\resizebox{7.cm}{!}{\includegraphics[bb= 59 370 308
548]{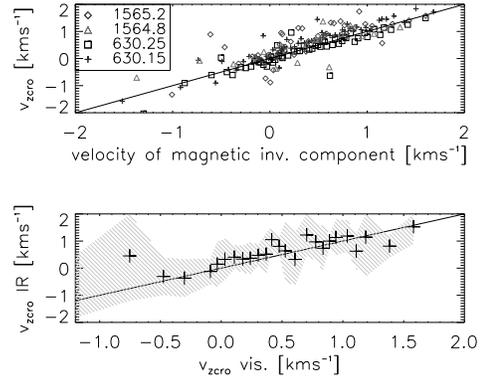}}}
\caption{{\em Top:} Stokes $V$ zero-crossing velocity, $v_{\rm zcro}$,
vs the magnetic velocity from the inversion, for BPs. {\em Bottom:} BP
zero-crossing velocities deduced from IR lines vs those from visible
lines. The zero-crossing velocity is averaged over the two visible
(or infrared) lines when both show regular Stokes $V$ profiles. The
shaded area indicates the scatter in the bins.\label{magvel}}
\end{figure}

The magnetic velocity indicated by the inversion usually agrees with the Stokes $V$ zero-crossing velocity deduced from the visible lines (upper panel of Fig.~\ref{magvel}), because a displacement between the observed and synthetic Stokes $V$ profiles would strongly degrade the quality of the fit, due to the steep slope of the circular polarization signal of the visible lines near the zero
crossing. The IR line at 1565.2 nm usually shows the strongest redshifts. \ion{Fe}{i} 1564.8 nm has less reliable statistics, as $v_{\rm zcro}$ is only calculated for regular profiles (Table \ref{proftypes}). The bottom panel of Fig.~\ref{magvel} compares the Stokes $V$
zero-crossing velocities derived from the IR and visible lines. The
plot also shows a trend toward higher velocities in the IR as compared
with the visible, as most points lie above the line of one-to-one
correlation. This difference in magnetic velocity could be due to unresolved
structures or flow fields, whose effect on infrared and visible lines
may differ. It is also compatible with the existence of
velocity gradients along the line of sight, as indicated by the
profile asymmetries discussed in the next section.
\begin{figure}
\centerline{\resizebox{6.cm}{!}{\includegraphics[bb= 57 365 283
505]{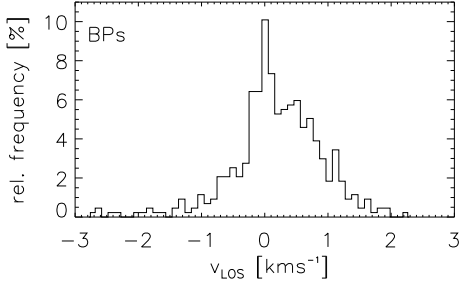}}}
\caption{Histogram of the LOS velocity of the magnetic component of
the inversion, for BPs.\label{intvel1}}
\end{figure}
\begin{figure*}
\centerline{\resizebox{14.2cm}{!}{\includegraphics[bb= 63 365 552
756]{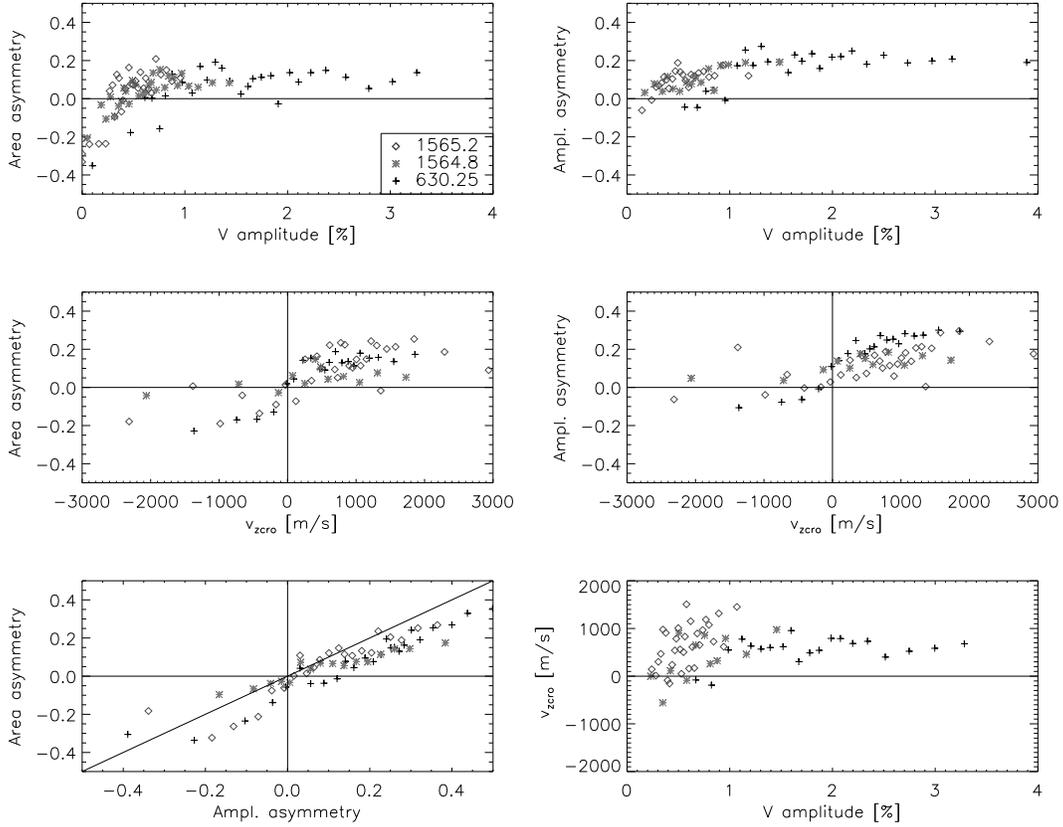}}}
\caption{Scatter plot of various Stokes $V$ line parameters for the BP sample. 
{\em Left to right, top to bottom:} $\delta A$ and $\delta a$ vs Stokes $V$ 
amplitude, $\delta A$ and $\delta a$ vs zero-crossing velocity, 
area asymmetry vs amplitude asymmetry, and zero-crossing velocity vs 
Stokes $V$ amplitude.\label{asymm}}
\end{figure*}

Table \ref{table_vel} summarizes the average velocities of BPs and
NBPs; positive values correspond to redshifts. The first and second row show the average zero-crossing velocity and the average line-core velocity of BPs for the different spectral lines. The third row contains the average line-core velocity of NPBs. In the fourth row, the convective blueshifts from the two-component quiet Sun model of \citet{borrero+bellot2002} are given that were used to set up the wavelength scale. For BPs, the average zero-crossing velocities are between 300~m
~s$^{-1}$ and 600~m~s$^{-1}$, depending on the spectral line; the
average velocity of the magnetic component of the inversion is
260~m~s$^{-1}$ (cf.~Fig.~\ref{intvel1}). These velocities
agree with the results of \citet{amer+kneer1993}, as far as
different velocities are deduced from Stokes $I$ and Stokes $V$, and
with \citet{grossmanndoerth+etal1996} or \citet{sigwarth+etal1999},
who found significant redshifts in magnetic elements. The line-core
velocity of the BPs shows blueshifts. The magnetic elements and
intergranular lanes are not resolved in the VTT data; thus, the
intensity profiles of the BPs are probably affected by unpolarized
light from bright granules in the immediate surroundings, which is a
source of blueshifts. For the NBP sample, we find average velocities
close to the convective blueshifts used in the determination of the
wavelength scale, as expected. As discussed in
Sect.~\ref{wavelengthscale}, the velocities may be biased towards the
blue.

\subsection{Area and amplitude asymmetries\label{assym}}
The area and amplitude asymmetries are sensitive to velocity and
magnetic field gradients along the line of
sight. \cite{auer+heasley1978} demonstrated that the existence of
velocity gradients along the LOS is a necessary and sufficient
condition for having non-zero area asymmetries
\citep[cf.~also][]{lopezariste2002}. Enhanced asymmetries result when
gradients of velocity are combined with gradients of magnetic field
strength, field inclination, and/or field azimuth. Note that jumps of
these atmospheric parameters at the interface between magnetic flux
concentrations and their field-free surroundings would effectively
produce gradients along the LOS. In the context of magnetic canopies,
\citet{solanki+pahlke1988}, \citet{almeida+iniesta1989}, or
\citet{almeida1998} have demonstrated that the sign of the area asymmetry
is related to the sign of the gradients of velocity and field strength
through
\begin{equation}
{\rm sgn}\, (\delta A) = - \, {\rm sgn} \left( \frac{{\rm
d}B(\tau)}{{\rm d}\tau} \frac{{\rm d}v(\tau)}{{\rm d}\tau}\right).
\end{equation}

In Fig.~\ref{asymm} we display the correlation between the asymmetries
and other parameters for the BPs. The upper panels show that most BPs
exhibit positive area and amplitude asymmetries, with values
comparable to those found by \citet{valentinetal1997} in plage regions
and by \citet{sigwarth+etal1999} in network and internetwork
areas. Negative asymmetries are encountered only for weak polarization
signals. Both $\delta A$ and $\delta a$ change sign with the zero-crossing
velocity (middle panels of Fig.~\ref{asymm}). As a consequence, there
are very few examples of, e.g., positive asymmetries associated with
blueshifted profiles. The lower left panel of Fig.~\ref{asymm} shows that the correlation
between area and amplitude asymmetry is quite tight for G-band BPs:
structures with positive area asymmetry also have positive amplitude
asymmetry, and when one increases the other follows it
closely. Another finding is that the zero-crossing velocity of BPs
depends on the amplitude of the polarization signal only for very low
signals: most BPs with strong Stokes $V$ signals show similar
velocities (cf.\ the lower right panel of Fig.~\ref{asymm}).
\begin{figure}[b]
\centerline{\resizebox{8.cm}{!}{\includegraphics[bb = 62 365 278
466]{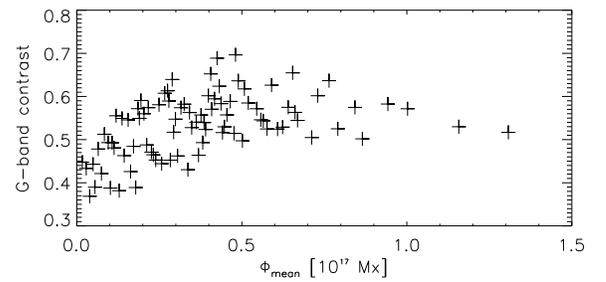}}}
\caption{G-band contrast vs average single-pixel magnetic flux of
BPs. The curve has a similar shape to the contrast vs total
integrated polarization (cf.\ Fig.~\ref{poldeg}).\label{invparam1mag}}
\end{figure}
\begin{figure}[b]
\includegraphics[bb= 167 700 425 782]{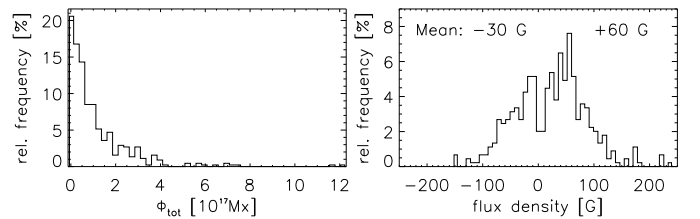}
\caption{{\em Left}: Histogram of the total unsigned flux of BPs. {\em
Right}: Maximum flux density of BPs. \label{fluxhist}}
\end{figure}
\begin{figure*}
\includegraphics{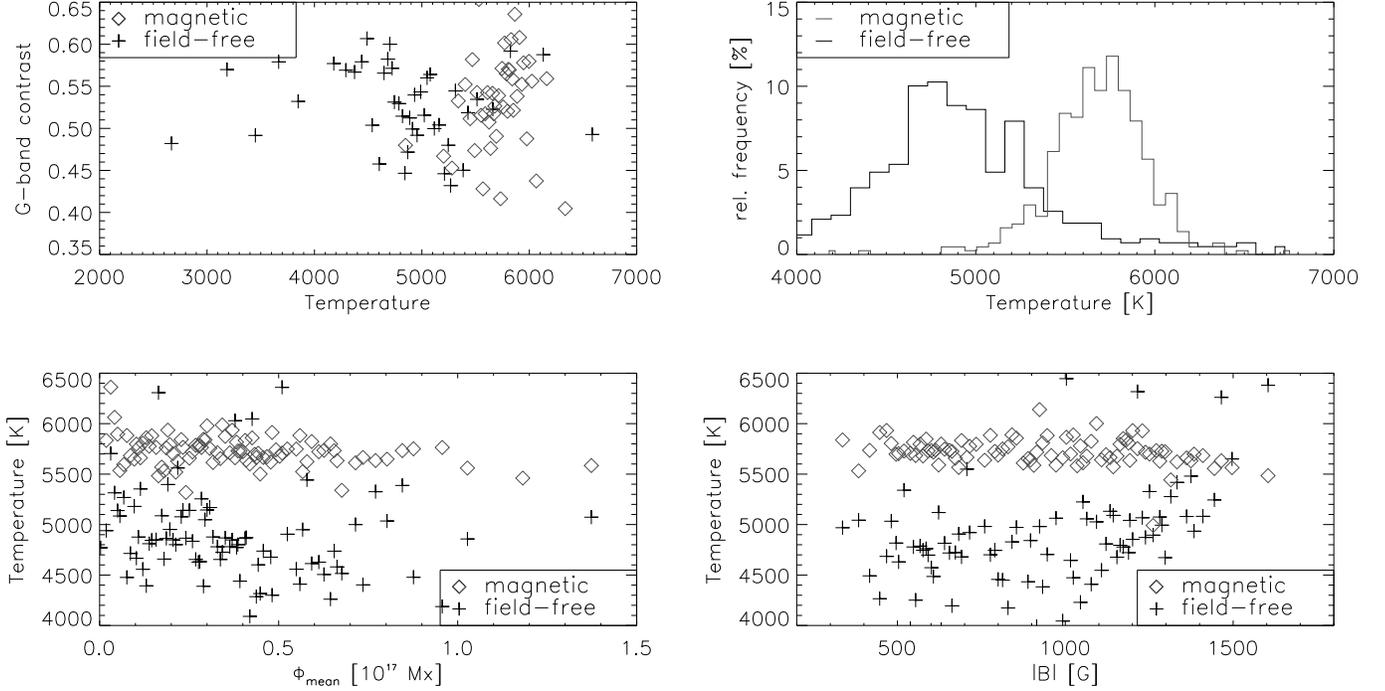}
\caption{{\em Top left:} G-band contrast vs temperature of the two
atmospheric components used to invert the BP profiles: the magnetic
interior ($\Diamond$) and the external, field-free surroundings
(+). {\em Top right:} Histogram of temperatures. The magnetic interior
is usually hotter than the surroundings by about 1000 K at the same
optical depth. {\em Bottom left:} Temperature of the two atmospheric
components vs average unsigned magnetic flux. The magnetic temperature
is seen to decrease linearly with flux.  {\em Bottom right:}
Temperature as a function of field strength.  A slight decrease in the
magnetic temperature is observed. For the field-free component, no
clear trends with $\Phi_{\rm mean}$ or $B$ exist.
\label{tempparam}\label{tempparam1}}
\end{figure*}

\subsection{Magnetic flux\label{magflux}}
The magnetic flux per pixel is calculated from the inversion results
as $\Phi = (1-\beta) \, f \, A \, B \, \cos \gamma$, where $A = 254^2$
km$^2$ is the area corresponding to a VTT pixel. Figure \ref{invparam1mag} displays the variation of the G-band
contrast of BPs with the average unsigned flux per pixel, $\langle\Phi\rangle_{\rm BP}$, where $\langle \rangle_{\rm BP}$ indicates the average over all VTT pixels covered by the BP. The contrast increases with magnetic flux until about $\langle\Phi\rangle_{\rm BP} = 5 \times 10^{16}$~Mx and
then decreases. The relation between G-band contrast and mean unsigned
flux is very similar to that between contrast and total integrated
polarization (${\cal T}$, cf.~Fig.~\ref{poldeg}). This supports
the consistency of the inversion results: ${\cal T}$ is proportional
to the flux and is derived directly from the observed profiles. The
magnetic flux is calculated using three different free parameters of
the inversion, but shows the same dependence with the G-band contrast.

The total magnetic fluxes we infer for the BPs, $\Phi_{\rm tot} = n \cdot \langle
\Phi\rangle_{\rm BP}$, with $n$ the number of VTT pixels covered by the BP, are similar to the ones obtained by
\citet{berger+title2001}, albeit generally smaller (see left panel of
Fig.~\ref{fluxhist}). For a comparison with their paper we also
computed the maximum value of the flux density in each BP ($\Phi_{\rm
max}({\rm BP})/A$; right panel of Fig.~\ref{fluxhist}), for which we
again find lower values than they do. The most probable value of
$\Phi_{\rm max}({\rm BP})/A$ is around $+60$~G for positive-polarity
BPs and $-30$~G for negative-polarity BPs, i.e., only about half the
value reported by \citet{berger+title2001}. This systematic difference
may have two origins: (1) the different methods used to derive the
magnetic flux, full vector polarimetry of several lines in the present
study and magnetograms of a single line in the case of
\citet{berger+title2001}, which result in different noise levels and
polarimetric sensitivities, and/or (2) the lower spatial resolution of
our observations.

\subsection{Dependence of G-band contrast on temperature}
The temperature stratification of the BPs is derived from the
inversion, both for the magnetic interior and the immediate field-free
surroundings. To facilitate comparisons, we consider the mean
temperature of each atmospheric component in the optical depth range
from $\log \tau = 0$ to $\log \tau = -2$. The relation between the G-band contrast and the temperature of BPs is
displayed in the upper left panel of Fig.~\ref{tempparam}. The
contrast shows no clear correlation with the temperature of either the
magnetic interior or the field-free surroundings. However, the
temperatures of the two components show systematic differences: the
histograms displayed in the upper right panel reveal that the magnetic
atmosphere is hotter than the immediate field-free surroundings by
about 1000 K at equal optical depth. The mean temperature is 5800 K
for the magnetic component and around 4800 K for the field-free
surroundings.
\subsection{Dependence of the temperature on magnetic flux and field strength}
\label{temp_flux}
A slight reduction in the temperature of the magnetic interior with
increasing magnetic flux is found, as can be seen in the lower left
panel of Fig.~\ref{tempparam}. A similar correlation exists between
the temperature of the magnetic component and the field strength
(lower right panel of Fig.~\ref{tempparam}). The temperature of the
field-free surroundings does not show any correlation with either the
magnetic flux or field strength.

\subsection{The size of G-band BPs}
\label{sizes}

The quantity $f_{\rm mag} = (1-\beta)\, f$ represents the fraction of
a single VTT pixel occupied by magnetic fields, as estimated by
the inversion code. It can be converted to the effective diameter $D$
of a circular structure with the same area. The upper limit of $D$ for
a single pixel is 292 km; however, some BPs cover $n$ VTT pixels. The
size of each BP in the VTT maps is thus calculated from $n \langle
f_{\rm mag}\rangle_{\rm BP}$. In this derivation it is implicitly
assumed that only a single source is generating the polarized light,
even if the magnetic filling fraction only gives the area inside the
pixel occupied by all fields. For the VTT, no lower size
limit exists, because the magnetic filling factor can go smoothly
towards zero. For the DOT images, the number of bright pixels
belonging to a single BP structure is converted to an effective
diameter. The minimum effective diameter for any BP in the DOT
filtergrams is then 60 km ($\equiv 1$ DOT pixel).

\begin{figure}
\centerline{\resizebox{8cm}{!}{\includegraphics{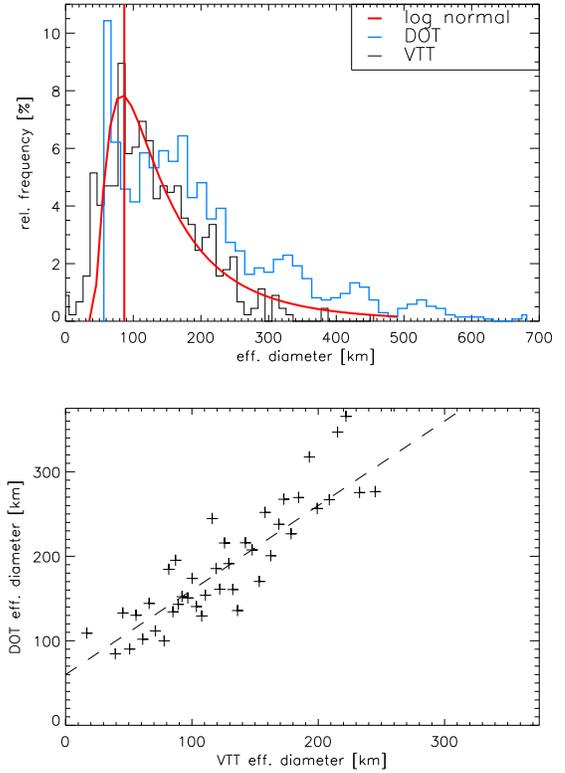}}}
\caption{{\em Top:} Histogram of BP effective diameters from the VTT
(thin black line) and the DOT (thick line). Overplotted is a
log-normal distribution. {\em Bottom:} Effective diameters from the
DOT vs those from the VTT. The dashed line has unity slope and 60~km
offset.\label{magfillfrac}}
\end{figure}

Figure \ref{magfillfrac} displays the histogram of BP diameters as
inferred from both data sets under these assumptions. The
maximum diameter is 500 km in the VTT maps and 700~km in the DOT
filtergrams. We find a distribution with a mean value of 150~km for
the VTT and 210~km for the DOT, with few BPs with diameters above
300~km. However, the mean value may be misleading due to the skewness
of the distribution. For comparison, the overplotted log-normal
distribution would indicate $D \sim 90$~km as the most 
probable BP diameter. In the DOT filtergrams, there is an increase of the relative frequency of small BP structures extending over 1 or 2 pixels (60-100~km). These
structures are very likely artifacts introduced by the data processing
and the alignment procedure of DOT and VTT data, where the DOT
image closest in time is used without a smooth transition from one DOT
image to the next. In principle, these smallest BPs should have been
rejected as unphysical, because the DOT resolution of 0\farcs2 forces
the minimum diameter to be around 170 km. Our manual removal of short
contour lines thus seems to have been insufficient, but fortunately
only a small fraction of the BPs is affected. Note that a more
sophisticated algorithm for the identification of BPs would also
most probably  have resolved the larger structures into chains of smaller
BPs. The BP sizes found here are comparable to the typical width of
intergranular lanes and the BP sizes given by \citet{berger+title2001}
and \citet{wiehr+etal2004}, respectively.

The lower panel of Fig.~\ref{magfillfrac} compares the effective
diameters of BPs inferred from the VTT and the DOT observations.  
The BP sizes derived from the former are systematically smaller,
amounting to only 78 \% of the DOT diameter on average. The smaller
size may be related to the spatial resolution of the VTT spectra. The
magnetic filling fraction is underestimated, as part of the polarized
light is scattered out of the VTT pixel by seeing.

\section{Summary and discussion\label{discussion}}

The analysis of cospatial and simultaneous G-band images and
spectropolarimetric measurements reveals different aspects of both the
statistics of BP fields and the relation between G-band intensities
and magnetic properties of the flux concentrations. We find the
following properties from a statistical analysis of 447 identified
BPs:

\begin{itemize}
\item 94\% of the BPs are cospatial with polarization signals above
the noise level.
\item The magnetic field strengths of BPs range from 500 to 1400~G. 
The field strength distribution is rather flat within this range. 
\item The total magnetic flux of BPs ranges from 0 to up to $4\times 10^{17}$~Mx, with an average flux density per pixel of around 50 G.
\item The distribution of magnetic field inclinations to the LOS, 
hence to the solar surface, indicates that not all BP fields are vertical. 
\item The BPs observed in a sunspot moat exhibit redshifts in the
magnetized plasma. The Stokes $V$ zero-crossing velocity derived from
the IR lines is around 500~m~s$^{-1}$, while that derived from the
visible lines is approximately 400~m~s$^{-1}$. The magnetic component
of the inversion shows a redshift of 260~m~s$^{-1}$ on
average. Moreover, the profiles of visible and IR lines exhibit
amplitude and area asymmetries that increase with the magnetic
velocity.
\item The magnetic interior of BPs is about 1000 K hotter than the
field-free surroundings at equal optical depth. The temperature of the
magnetic component is slightly reduced for large magnetic fluxes.
\item The area covered by individual BPs in the DOT filtergrams is roughly 
consistent with the size of BPs inferred from the inversion of the
visible and infrared lines. The average effective diameter of BPs is around 100 to 150 km, with few structures larger than 300 km.
\end{itemize}

For the relations between G-band intensity and magnetic properties 
of the flux concentrations, we find that:
\begin{itemize}
\item The G-band contrast of BPs increases slightly with the magnetic field
strength.
\item The contrast only slightly depends on the magnetic flux per
pixel. It increases with flux for values below $5 \times 10^{16}$~Mx
and decreases with flux for values above $10^{17}$~Mx.
\item The G-band contrast of BPs scales with the inclination: the 
smaller the angle between the magnetic field and the line of sight, 
the higher the G-band contrast.
\end{itemize}
\begin{figure}
\centerline{\resizebox{6.5cm}{!}{\includegraphics{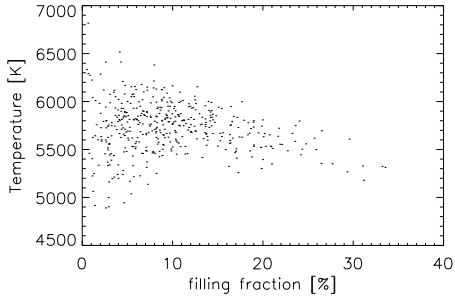}}}
\caption{Temperature of the magnetic atmosphere vs its filling 
fraction. For large filling factors, a reduction of the temperature 
is seen.\label{filltemp}}
\end{figure}

Most of these observational findings agree with the
expected properties of small-scale magnetic flux elements. First, 
magnetic elements in horizontal temperature equilibrium have lower 
gas densities, in order to maintain lateral pressure balance with the 
external field-free medium. Due to the reduced gas density,
a given optical depth corresponds to lower geometrical heights in the
magnetic interior than in the field-free surroundings: inside the
magnetic flux concentration, deeper and hence hotter layers are
seen. The downward shift of the optical depth scale depends on the
magnetic field strength, which enters the pressure balance equation
\citep{spruit1976,steiner+stenflo1990,bellot+cobo+collados2000,
schuessler+etal2003,shelyag+etal2004}. Second, a G-band BP 
essentially outlines a deficit in the abundance of the CH
molecule. Under thermodynamic equilibrium conditions, the amount of CH
in the solar atmosphere depends both on the gas density and on the
temperature \citep{jorgeetal2001,steiner+hauschildt+bruls2001,
langhans+schmidt+tritschler2002}. A strong enhancement of the 
G-band intensity contrast with respect to nearby continuum 
wavelengths indicates a weakening of the CH lines 
due to increased thermal dissociation.

We want to explicitly point out the relation between some of our 
observational findings and the modeling of flux concentrations. 
The temperature difference between the magnetic and field-free 
components of the inversion at the same optical depth (Fig.\ 18) 
agrees with a shift of the optical depth scale in the presence 
of magnetic fields. The G-band contrast is found to increase
by around 0.2 for fields from 0.4 to 1.5 kG, suggesting a stronger
shift of the optical depth scale for higher field strengths. However,
the trend falls short of the predictions from simulations
\citep[e.g.][their Fig.\ 4]{shelyag+etal2004}, which suggest 
a contrast increase greater than 0.5 over the same range.

The slight decrease in the G-band contrast for magnetic fluxes larger 
than $10^{17}$~Mx is at first not predicted by models of small flux
elements. However, we suggest that it is actually related to a
temperature effect. Since there is an upper limit to the field
strength on the order of 1.5 kG (cf.~Fig.~\ref{histfldstrg}), the
amount of flux must be proportional to the diameter of the flux
concentration. The volume inside the flux element to be heated
increases faster with the diameter of the flux concentration 
($\propto r^2 dh$) than does the interface with its surroundings 
($\propto r dh$). Thus, larger magnetic flux should lead to larger areas, 
 hence to reduced temperatures \citep{spruit1976}. While 
this argument is rather indirect, the inversion results 
allow us to directly check the relation between temperature 
and area. Figure \ref{filltemp} displays a scatter plot 
of the temperature of the magnetic component of the inversion 
vs the magnetic filling fraction, which is proportional to 
the area. In this plot, we see a clear trend toward a reduction 
of the temperature with filling factor, hence with area.

\begin{figure}
\centerline{\resizebox{3.2cm}{!}{\includegraphics{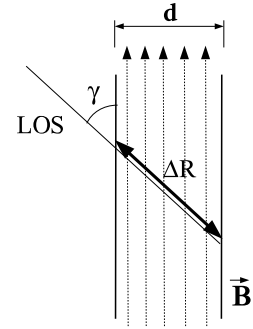}}}
\caption{Sketch of an isolated flux concentration embedded in field-free
surroundings. The magnetic field is along the vertical direction and the LOS
is inclined by an angle $\gamma$ with respect to it. The path length of a ray
inside the flux concentration, $\Delta R$, depends on both $\gamma$ and the 
thickness $d$ of the tube. \label{fluxgeometry}}
\end{figure}

Perhaps the most surprising result of our analysis is the dependence
of the G-band contrast on the magnetic field inclination to the LOS
displayed in the middle panels of Fig.~\ref{lpv}. We note that the
dependence seems to have the form
\begin{equation}
\label{empirical}
{\cal C} = a + \frac{b}{\sin \gamma}, 
\end{equation}
where $\gamma$ represents the field inclination and $a$ and $b$ are
constant parameters. A very simple model of a vertical flux
concentration embedded in field-free surroundings
(Fig.~\ref{fluxgeometry}) shows that the second term on the right hand
side of Eq.~(\ref{empirical}) is proportional to $\Delta R$, the
distance inside the flux concentration traveled by a ray that hits the
magnetic element at an angle $\gamma$. In fact, $\Delta R = d/\sin
\gamma$. The absorption along the LOS would then be $\propto \Delta R
\cdot \rho(B)$, with the density $\rho(B)$ inside the flux
concentration being lower than the density outside it. Thus, $\Delta
R$ would finally translate into a downward shift of the optical depth
scale for lines of sight that pass through the magnetic element,
compared with others that miss it. Note that $\Delta R$ is limited by
a $\Delta R_{\rm max}$, the depth where the atmosphere gets opaque
even for $\gamma = 0$ deg. We have overplotted three curves for
different values of a and b in the middle panels of Fig.~\ref{lpv},
where the middle curve ($b$ = 200)  at least roughly agrees with
the observed inclinations and L/V-values.

\section{Conclusions}
\label{summary}
We have derived the thermal, magnetic, and kinematic properties of
G-band bright points in the moat of a regular sunspot from
infrared and visible spectral lines observed at the German Vacuum
Tower Telescope on Tenerife. The BPs were identified in cospatial
diffraction-limited filtergrams taken with the Dutch Open Telescope on
La Palma. The Stokes profiles of the infrared and visible spectral
lines were inverted simultaneously using a two-component model
atmosphere with height-independent magnetic fields and line-of-sight 
(LOS) velocities. In addition, line parameters were extracted
from the observed Stokes profiles for an estimate of field strengths,
field inclinations, and LOS velocities.

We conclude that G-band brightenings are caused by concentrated
magnetic fields in more than $90\%$ of the cases. The minimum
requirement seems to be a field strength of at least 500~G. However,
the G-band BPs show a variety of magnetic and kinematic properties, as
suggested by the broad range of values we find in all physical
quantities ($B$, $\gamma$, $\Phi$, etc). Several patches of magnetic
flux can be traced in the polarimetric data during the full
1-hour time series, which show up as BP only part of the time. Together
with the broad range of field strengths, we think this indicates that
{\em not all BPs are cospatial with stable kG flux tubes}. We find a
stronger dependence of the G-band brightness on a geometrical effect,
namely on the inclination of the magnetic field to the LOS, than on most
of the other physical quantities. We conclude that for an accurate
description of the generation of G-band BPs or, more generally, 
the intensity in the G band, it is necessary to develop models of 
flux concentrations and their surroundings in at least two dimensions, 
to take into account the geometry of the field, its strength, 
and the actual viewing angle of the observations.

Even if our data corresponds to the moat of a sunspot, we believe that 
the magnetic properties of the observed BPs are not strongly influenced 
by the presence of the spot. The distance to the spot boundary is 
several arcseconds in most cases. Thus, the results of this paper 
might  also apply to BPs observed in other active and non-active 
regions. 

Finally, we stress that the use of G-band brightness enhancements 
as proxies of magnetic fields may miss part of the solar magnetic 
flux, as fields with strengths below 500~G remain undetected.

\begin{acknowledgements}
Discussions with O.\ Steiner and A.\ Tritschler are gratefully
acknowledged.  This work has been supported by the German DFG under
grant SCHL 514/2--1 and by the Spanish MEC under {\em Programa
Ram\'on y Cajal} and project ESP2003-07735-C04-03. The DOT is operated
by Utrecht University at the Spanish Observatorio del Roque de los
Muchachos of the Instituto de Astrof\'{\i}sica de Canarias (IAC). The
VTT is operated by the Kiepenheuer-Institut f\"ur Sonnenphysik at the
Spanish Observatorio del Teide, also of the IAC. We thank the referee, 
Dr.~J.~S{\'a}nchez Almeid{\'a}, for his suggestions for improving the paper.

\end{acknowledgements}

\bibliographystyle{aa} 
\bibliography{5620_ref} 

\begin{thebibliography}{49}
\expandafter\ifx\csname natexlab\endcsname\relax\def\natexlab#1{#1}\fi

\bibitem[{{Amer} \& {Kneer}(1993)}]{amer+kneer1993}
{Amer}, M.~A. \& {Kneer}, F. 1993, \aap, 273, 304

\bibitem[{{Auer} \& {Heasley}(1978)}]{auer+heasley1978}
{Auer}, L.~H. \& {Heasley}, J.~N. 1978, \aap, 64, 67

\bibitem[{{Ballesteros} {et~al.}(1996){Ballesteros}, {Collados}, {Bonet},
  {Lorenzo}, {Viera}, {Reyes}, \& {Rodriguez Hidalgo}}]{ballesteros+etal1996}
{Ballesteros}, E., {Collados}, M., {Bonet}, J.~A., {et~al.} 1996, \aaps, 115,
  353

\bibitem[{{Balthasar} {et~al.}(1996){Balthasar}, {Schleicher}, {Bendlin}, \&
  {Volkmer}}]{balthasar+etal1996}
{Balthasar}, H., {Schleicher}, H., {Bendlin}, C., \& {Volkmer}, R. 1996, \aap,
  315, 603

\bibitem[{{Beck} {et~al.}(2005{\natexlab{a}}){Beck}, {Schlichenmaier},
  {Collados}, {Bellot Rubio}, \& {Kentischer}}]{beck+etal2005a}
{Beck}, C., {Schlichenmaier}, R., {Collados}, M., {Bellot Rubio}, L., \&
  {Kentischer}, T. 2005{\natexlab{a}}, \aap, 443, 1047

\bibitem[{{Beck} {et~al.}(2005{\natexlab{b}}){Beck}, {Schmidt}, {Kentischer},
  \& {Elmore}}]{beck+etal2005b}
{Beck}, C., {Schmidt}, W., {Kentischer}, T., \& {Elmore}, D.
  2005{\natexlab{b}}, \aap, 437, 1159

\bibitem[{{Bellot Rubio} {et~al.}(2000){Bellot Rubio}, {Ruiz Cobo}, \&
  {Collados}}]{bellot+cobo+collados2000}
{Bellot Rubio}, L.~R., {Ruiz Cobo}, B., \& {Collados}, M. 2000, \apj, 535, 489

\bibitem[{{Berger} \& {Title}(1996)}]{berger+title1996}
{Berger}, T.~E. \& {Title}, A.~M. 1996, \apj, 463, 365

\bibitem[{{Berger} \& {Title}(2001)}]{berger+title2001}
{Berger}, T.~E. \& {Title}, A.~M. 2001, \apj, 553, 449

\bibitem[{{Bonet} {et~al.}(2005){Bonet}, {M{\'a}rquez}, {Muller}, {Sobotka}, \&
  {Roudier}}]{bonetetal2004}
{Bonet}, J.~A., {M{\'a}rquez}, I., {Muller}, R., {Sobotka}, M., \& {Roudier},
  T. 2005, \aap, 430, 1089

\bibitem[{{Borrero} \& {Bellot Rubio}(2002)}]{borrero+bellot2002}
{Borrero}, J.~M. \& {Bellot Rubio}, L.~R. 2002, \aap, 385, 1056

\bibitem[{{Bovelet} \& {Wiehr}(2003)}]{bovelet+wiehr2003}
{Bovelet}, B. \& {Wiehr}, E. 2003, \aap, 412, 249

\bibitem[{{Carlsson} {et~al.}(2004){Carlsson}, {Stein}, {Nordlund}, \&
  {Scharmer}}]{carlsson+etal2004}
{Carlsson}, M., {Stein}, R.~F., {Nordlund}, {\AA}., \& {Scharmer}, G.~B. 2004,
  \apjl, 610, L137

\bibitem[{{Dunn} \& {Zirker}(1973)}]{dunn+zirker1973}
{Dunn}, R.~B. \& {Zirker}, J.~B. 1973, \solphys, 33, 281

\bibitem[{{Gingerich} {et~al.}(1971){Gingerich}, {Noyes}, {Kalkofen}, \&
  {Cuny}}]{gingerich+etal1971}
{Gingerich}, O., {Noyes}, R.~W., {Kalkofen}, W., \& {Cuny}, Y. 1971, \solphys,
  18, 347

\bibitem[{{Grossmann-Doerth} {et~al.}(1996){Grossmann-Doerth}, {Keller}, \&
  {Schuessler}}]{grossmanndoerth+etal1996}
{Grossmann-Doerth}, U., {Keller}, C.~U., \& {Schuessler}, M. 1996, \aap, 315,
  610

\bibitem[{{Keller}(1992)}]{keller1992}
{Keller}, C.~U. 1992, \nat, 359, 307

\bibitem[{{Khomenko} {et~al.}(2003){Khomenko}, {Collados}, {Solanki}, {Lagg},
  \& {Trujillo Bueno}}]{khomenko+etal2003}
{Khomenko}, E.~V., {Collados}, M., {Solanki}, S.~K., {Lagg}, A., \& {Trujillo
  Bueno}, J. 2003, \aap, 408, 1115

\bibitem[{{L{\' o}pez Ariste}(2002)}]{lopezariste2002}
{L{\' o}pez Ariste}, A. 2002, \apj, 564, 379

\bibitem[{{Landi Degl'Innocenti}(2003)}]{landi2003}
{Landi Degl'Innocenti}, E. 2003, AN, 324, 393

\bibitem[{{Langhans} {et~al.}(2004){Langhans}, {Schmidt}, \&
  {Rimmele}}]{langhans+etal2004}
{Langhans}, K., {Schmidt}, W., \& {Rimmele}, T. 2004, \aap, 423, 1147

\bibitem[{{Langhans} {et~al.}(2002){Langhans}, {Schmidt}, \&
  {Tritschler}}]{langhans+schmidt+tritschler2002}
{Langhans}, K., {Schmidt}, W., \& {Tritschler}, A. 2002, \aap, 394, 1069

\bibitem[{{Lites}(2002)}]{lites2002}
{Lites}, B.~W. 2002, \apj, 573, 431

\bibitem[{{Mart{\' i}nez Pillet} {et~al.}(1999){Mart{\' i}nez Pillet},
  {Collados}, {S{\' a}nchez Almeida}, {Gonz{\' a}lez}, {Cruz-Lopez},
  {Manescau}, {Joven}, {Paez}, {Diaz}, {Feeney}, {S{\' a}nchez}, {Scharmer}, \&
  {Soltau}}]{martinez+etal1999}
{Mart{\' i}nez Pillet}, V., {Collados}, M., {S{\' a}nchez Almeida}, J.,
  {et~al.} 1999, in ASP Conf. Ser. 183: High Resolution Solar Physics: Theory,
  Observations, and Techniques, 264

\bibitem[{{Mart\'{\i}nez Pillet} {et~al.}(1997){Mart\'{\i}nez Pillet}, {Lites},
  \& {Skumanich}}]{valentinetal1997}
{Mart\'{\i}nez Pillet}, V., {Lites}, B., \& {Skumanich}, A. 1997, \apj, 474,
  810

\bibitem[{{Mehltretter}(1974)}]{mehltretter1974}
{Mehltretter}, J.~P. 1974, \solphys, 38, 43

\bibitem[{{Muller}(1983)}]{muller1983}
{Muller}, R. 1983, \solphys, 85, 113

\bibitem[{{Press} {et~al.}(1986){Press}, {Flannery}, \&
  {Teukolsky}}]{press+etal1986}
{Press}, W.~H., {Flannery}, B.~P., \& {Teukolsky}, S.~A. 1986, {Numerical
  recipes. The art of scientific computing} (Cambridge: University Press, 1986)

\bibitem[{{Reardon}(2006)}]{reardon2006}
{Reardon}, K.~P. 2006, \solphys, 239, 503

\bibitem[{{Rezaei} {et~al.}(2006){Rezaei}, {Schlichenmaier}, {Beck}, \& {Bellot
  Rubio}}]{reza+etal2006}
{Rezaei}, R., {Schlichenmaier}, R., {Beck}, C., \& {Bellot Rubio}, L.~R. 2006,
  \aap, 454, 975

\bibitem[{{Ruiz Cobo} \& {del Toro Iniesta}(1992)}]{cobo+toroiniesta1992}
{Ruiz Cobo}, B. \& {del Toro Iniesta}, J.~C. 1992, \apj, 398, 375

\bibitem[{{S{\" u}tterlin} {et~al.}(2004){S{\" u}tterlin}, {Bellot Rubio}, \&
  {Schlichenmaier}}]{suetterlin+etal2004}
{S{\" u}tterlin}, P., {Bellot Rubio}, L.~R., \& {Schlichenmaier}, R. 2004,
  \aap, 424, 1049

\bibitem[{{Sanchez Almeida}(1998)}]{almeida1998}
{Sanchez Almeida}, J. 1998, \apj, 497, 967

\bibitem[{{S{\'a}nchez Almeida} {et~al.}(2001{\natexlab{a}}){S{\'a}nchez
  Almeida}, {Asensio Ramos}, {Trujillo Bueno}, \&
  {Cernicharo}}]{almeida+etal2001}
{S{\'a}nchez Almeida}, J., {Asensio Ramos}, A., {Trujillo Bueno}, J., \&
  {Cernicharo}, J. 2001{\natexlab{a}}, \apj, 555, 978

\bibitem[{{S{\'a}nchez Almeida} {et~al.}(2001{\natexlab{b}}){S{\'a}nchez
  Almeida}, {Asensio Ramos}, {Trujillo Bueno}, \& {Cernicharo}}]{jorgeetal2001}
{S{\'a}nchez Almeida}, J., {Asensio Ramos}, A., {Trujillo Bueno}, J., \&
  {Cernicharo}, J. 2001{\natexlab{b}}, \apj, 555, 978

\bibitem[{{Sanchez Almeida} {et~al.}(1989){Sanchez Almeida}, {Collados}, \&
  {del Toro Iniesta}}]{almeida+iniesta1989}
{Sanchez Almeida}, J., {Collados}, M., \& {del Toro Iniesta}, J.~C. 1989, \aap,
  222, 311

\bibitem[{{Sch{\" u}ssler} {et~al.}(2003){Sch{\" u}ssler}, {Shelyag},
  {Berdyugina}, {V{\" o}gler}, \& {Solanki}}]{schuessler+etal2003}
{Sch{\" u}ssler}, M., {Shelyag}, S., {Berdyugina}, S., {V{\" o}gler}, A., \&
  {Solanki}, S.~K. 2003, \apjl, 597, L173

\bibitem[{{Schmidt} {et~al.}(2003){Schmidt}, {Beck}, {Kentischer}, {Elmore}, \&
  {Lites}}]{schmidt+etal2003}
{Schmidt}, W., {Beck}, C., {Kentischer}, T., {Elmore}, D., \& {Lites}, B. 2003,
  Astronomische Nachrichten, 324, 300

\bibitem[{{Schmidt} \& {Kentischer}(1995)}]{schmidt+kentischer1995}
{Schmidt}, W. \& {Kentischer}, T. 1995, \aaps, 113, 363

\bibitem[{{Shelyag} {et~al.}(2004){Shelyag}, {Sch{\"u}ssler}, {Solanki},
  {Berdyugina}, \& {V{\"o}gler}}]{shelyag+etal2004}
{Shelyag}, S., {Sch{\"u}ssler}, M., {Solanki}, S.~K., {Berdyugina}, S.~V., \&
  {V{\"o}gler}, A. 2004, \aap, 427, 335

\bibitem[{{Sigwarth} {et~al.}(1999){Sigwarth}, {Balasubramaniam}, {Kn{\"
  o}lker}, \& {Schmidt}}]{sigwarth+etal1999}
{Sigwarth}, M., {Balasubramaniam}, K.~S., {Kn{\" o}lker}, M., \& {Schmidt}, W.
  1999, \aap, 349, 941

\bibitem[{{Solanki} \& {Pahlke}(1988)}]{solanki+pahlke1988}
{Solanki}, S.~K. \& {Pahlke}, K.~D. 1988, \aap, 201, 143

\bibitem[{{Spruit}(1976)}]{spruit1976}
{Spruit}, H.~C. 1976, \solphys, 50, 269

\bibitem[{{Steiner} {et~al.}(2001){Steiner}, {Hauschildt}, \&
  {Bruls}}]{steiner+hauschildt+bruls2001}
{Steiner}, O., {Hauschildt}, P.~H., \& {Bruls}, J. 2001, \aap, 372, L13

\bibitem[{{Steiner} \& {Stenflo}(1990)}]{steiner+stenflo1990}
{Steiner}, O. \& {Stenflo}, J.~O. 1990, in IAU Symposium, 181

\bibitem[{{Uitenbroek} \& {Tritschler}(2006)}]{uitenbroek+tritschler2006}
{Uitenbroek}, H. \& {Tritschler}, A. 2006, \apj, 639, 525

\bibitem[{{van Ballegooijen} {et~al.}(1998){van Ballegooijen}, {Nisenson},
  {Noyes}, {L{\" o}fdahl}, {Stein}, {Nordlund}, \&
  {Krishnakumar}}]{vanBallegooijen+etal1998}
{van Ballegooijen}, A.~A., {Nisenson}, P., {Noyes}, R.~W., {et~al.} 1998, \apj,
  509, 435

\bibitem[{{Wiehr} {et~al.}(2004){Wiehr}, {Bovelet}, \&
  {Hirzberger}}]{wiehr+etal2004}
{Wiehr}, E., {Bovelet}, B., \& {Hirzberger}, J. 2004, \aap, 422, L63

\bibitem[{{Zakharov} {et~al.}(2005){Zakharov}, {Gandorfer}, {Solanki}, \&
  {L{\"o}fdahl}}]{zakharovetal2005}
{Zakharov}, V., {Gandorfer}, A., {Solanki}, S.~K., \& {L{\"o}fdahl}, M. 2005,
  \aap, 437, L43

\end{thebibliography}
\clearpage
\newpage
\begin{appendix}
\section{Spatial resolution\label{spatres}}
\begin{figure}
\includegraphics{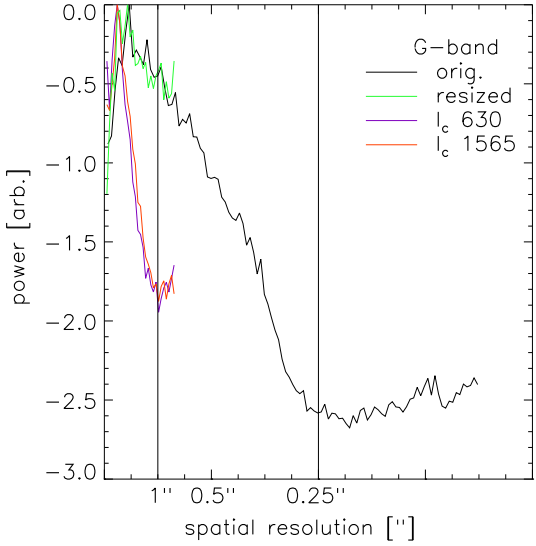}
\caption{Power spectra of the different data sets of the 1st scan as function of spatial resolution.\label{figspatres}}
\end{figure}
For an estimate of the spatial resolution, we calculated the Fourier transform of the intensity maps of the first scan across the sunspot. We then integrated the power of the Fourier transform over rings corresponding to increasing spatial frequencies. The inverse of the spatial frequency corresponds to a spatial scale. The resolution limit was estimated as the point, where the power spectrum levels off into the constant noise contribution. For the G-band data from the DOT, we find a resolution of around 0\farcs25 for the original  DOT map with 0\farcs071 square pixels (cf.~Fig.~\ref{figspatres}). When the data is resized to the VTT resolution of 0\farcs37 square pixels, the power is still conserved. For the polarimetric data in either infrared or visible continuum, we find a resolution of 1\farcs0 with this method.

\section{Alignment of data sets \label{alignsect}}
The alignment procedure uses the TIP pixel as a reference system:
for each position $(x,y)_{\rm TIP}$ in the TIP map, where $x$ corresponds to
the scan step and $y$ to the position along the slit, the algorithm looks for
cospatial points in the visible and UV channel of POLIS and the DOT
filtergrams, respectively. The TIP data have a spatial sampling of 0\farcs35
per pixel.  The data from the visible channel of POLIS have a finer sampling
of 0\farcs145, which is degraded to the TIP resolution by the linear
interpolation scheme described in the next section. The data from the VTT show displacements due to the actual position of the TIP and POLIS cameras and an additional variable offset caused by differential refraction \citep[e.g.][]{reardon2006} in the Earth's atmosphere.

\begin{figure}
\centerline{\resizebox{8cm}{!}{\includegraphics{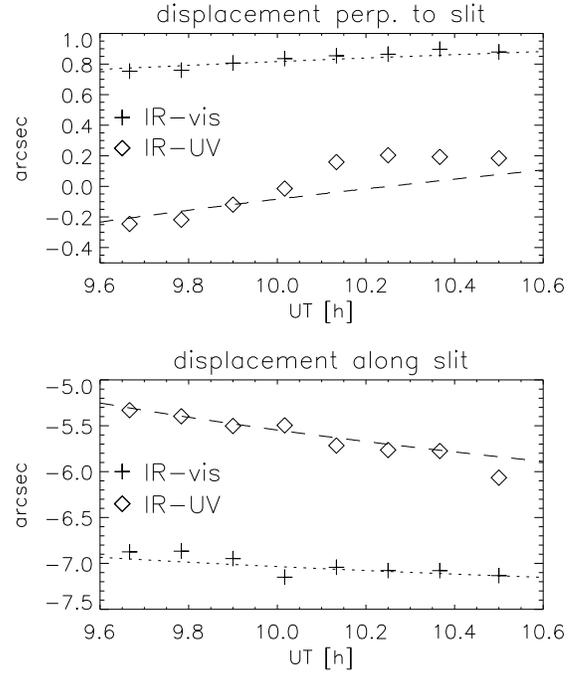}}}
\caption{Displacements between the visible and infrared intensity maps
($+$), and between the UV and infrared intensity maps ($\diamond$), in
arcsec.  For each repetition of the scan across the spot displayed in
Fig.~\ref{coarseali}, the shifts perpendicular to the slit ($\Delta
x$; {\em top}) and along the slit ($\Delta y$; {\em bottom}) have been
calculated through cross correlation. The dotted and dashed lines give
the shift values used for the alignment of each scan
step.\label{vttali}}
\end{figure}

\subsection{Alignment between IR and visible/UV data}
\paragraph*{Calculation of shifts.}
To find cospatial points in the TIP and POLIS data, the continuum intensity
maps of each repetition of the scan are correlated separately. For the UV
channel, a map of the line wing intensity of \ion{Ca}{ii} H is used, as it
shows photospheric structure (cf.\ Fig.~\ref{coarseali}).  The resulting shift
values along and perpendicular to the slit for the different times of
the eight repetitions allow  a curve to be fitted to the displacements 
(Fig.~\ref{vttali}). The best-fit curve is used to determine the 
shifts in $x$ and $y$ needed to align the TIP and POLIS observations 
for each scan step (taken every 6 seconds).

\paragraph*{Application of shifts. }
The spatial sampling of 0\farcs29 in the UV channel of POLIS is 
similar to that of TIP. For this reason, the shift 
$(\Delta x_{UV}, \Delta y_{UV})$ is simply added to the position 
$(x,y)_{\rm TIP}$, and the closest pixel is taken: 
\begin{equation}
(x,y)_{\rm UV} = {\rm round} \left[ (x,y)_{\rm TIP}+ (\Delta x_{\rm UV}, \Delta
y_{\rm UV}) \right] \;,
\end{equation}
where the rounding automatically selects the nearest cospatial pixel.

The spatial sampling in the visible channel of POLIS is almost three times
better than that of TIP. The cospatial Stokes profiles are then retrieved 
as a weighted average of the visible profiles along and perpendicular to
the slit. First, the cospatial position $(x,y)_{\rm vis}$ is given by
\begin{equation}
(x,y)_{\rm vis} = (x,y)_{\rm TIP}+ ({\rm floor}[\Delta x_{\rm vis}],
\Delta y_{\rm vis}),
\end{equation}
where ${\rm floor} [\Delta x_{\rm vis}]$ returns the largest 
integer smaller than $x_{\rm vis}$.

\begin{figure}
\centerline{\resizebox{8.5cm}{!}{\includegraphics{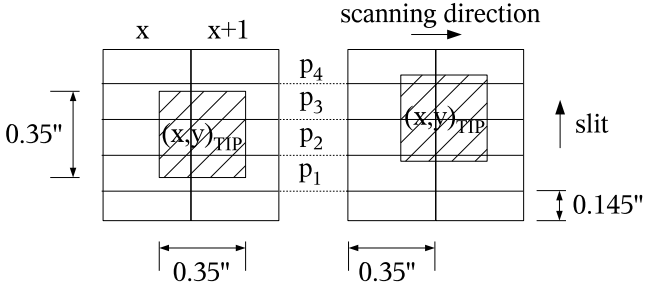}}}
\caption{Calculation of the visible profile that is cospatial to the 
infrared Stokes profile at $(x,y)_{\rm TIP}$. Two cases are possible: the 
TIP pixel covers three ({\em left}) or four ({\em right}) 
POLIS pixels along the slit.  The profiles $p_i$ from 
the two scan steps $x_{\rm vis}$ and $x_{\rm vis}+1$ are averaged 
with weights corresponding to the fraction of the POLIS resolution 
elements covered.\label{interpolscheme}}
\end{figure}

The TIP pixel partly covers several POLIS pixels
from two different scan steps, $x_{\rm vis}$ and $x_{\rm vis}+1$
(Fig.~\ref{interpolscheme}). The cospatial Stokes profile $p(\lambda,x,y)_{\rm
vis}$ is calculated as the weighted average
\begin{equation}
p(\lambda,x,y)_{\rm vis} = \frac{\sum_i a_i p(\lambda,x,y_i)_{\rm
vis}}{\sum_i a_i} + \frac{\sum_i b_i p(\lambda,x+1,y_i)_{\rm
vis}}{\sum_i b_i} \;,
\end{equation}
where $i= 1,\ldots,3$ or $i=1,\ldots,4$. The weights $a_i$ and $b_i$ are 
set according to the fraction of the POLIS pixels covered by 
the TIP pixel and verify the condition $\sum_i (a_i+b_i)=1$.

\subsection{Alignment between DOT and VTT data\label{dotalignment}}
In this case, cospatial {\em and} cotemporal positions have to be found. 
A cotemporal map from the DOT time series (``pseudo-scan'') is first 
constructed and then spatially aligned with the VTT data.

\subsubsection{Cotemporal DOT map}
The DOT pseudo-scan is the image that would result from stepping the 
POLIS/TIP slit across the DOT FOV. To create this map, the POLIS slit-jaw 
images are used. They display a 100\arcsec $\times$ 100\arcsec\, FOV of the 
solar surface centered on the slit. Each slit-jaw image is correlated with 
the DOT blue continuum filtergram closest in time, degraded to the same 
spatial resolution. From the correlation one finds the shifts in $x$ and 
$y$  required to align the images from the two telescopes. The POLIS 
slit visible in the slit-jaw image is superimposed on the coaligned DOT 
filtergram.  A slice of 5 pixel width ($= 0\farcs355$) is taken from 
the DOT images at the slit position and placed accordingly in the 
pseudo-scan map. Examples of the resulting pseudo-scan maps for the 
three DOT channels are displayed in Fig.~\ref{coarseali}.

\subsubsection{Cospatial DOT map}
\begin{figure}
\centerline{\resizebox{7.5cm}{!}{\includegraphics{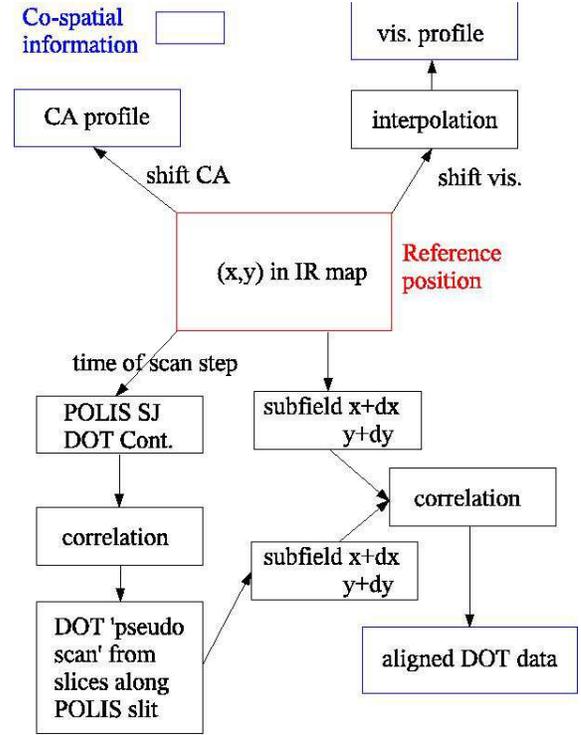}}}
\caption{Alignment procedure. For each position $(x,y)_{\rm TIP}$ in the TIP
data set, cospatial POLIS profiles are found by applying appropriate shifts 
to the visible and UV channels. To align the DOT filtergrams, the POLIS
slit-jaw image and the DOT continuum image closest in time are correlated and
used for the creation of a {\em pseudo-scan} map.  The final DOT/VTT alignment 
is done by taking a subfield of the TIP map around $(x,y)_{\rm TIP}$, the
corresponding subfield from the DOT pseudo-scan, and then correlating the
subfields.\label{flowchart}}
\end{figure}

The final alignment of the DOT images is achieved by taking a 25$\times$25
pixel subfield of the TIP map around each pixel $(x,y)_{\rm
TIP}$, the corresponding area from the DOT pseudo-scan continuum map, and
correlating the two subfields.  This procedure yields the position of a
5$\times$5 pixel area in the DOT maps that is cospatial with the TIP
pixel. The area is cut out from the three DOT pseudo-scan maps
and placed in new maps accordingly. Fig.~\ref{finalali} displays the coaligned
maps for the first repetition of the scan. The flow chart displayed in Fig.~\ref{flowchart} summarizes all the steps we follow to align the various data sets. 
\section{Coaligned maps\label{aligneddata}}
\begin{figure*}[t]
\caption{Coaligned maps from the VTT and the DOT for the first four
repetitions of the scan. {\em Clockwise, starting from top left of
each $4 \times 4$ subpanel:} G-band intensity, total integrated polarization, 
DOT BP mask, and IR continuum intensity. Blue contours outline enhanced
polarization signals and red contours the selected BP areas. In the
DOT BP mask, the BPs are color-coded according to their number. 
The lower right map of the first $4 \times4 $ subpanel is the VTT BP mask, 
not the DOT BP mask. In this map, the uniform pink area marks the 
spatial points used to extract the NBP sample. They are located 
outside the canopy, and at least 3 pixels (1\farcs05) away from any BP. 
 \label{alimapss1}}
\end{figure*}

Coaligned DOT and VTT maps are displayed in Figs.~\ref{alimapss1} and
\ref{alimapss2} for the eight repetitions of the scan. We show the aligned DOT G-band, the total integrated polarization, the IR continuum
intensity, and the DOT BP mask. For the first repetition of the scan,
the VTT mask of identified BPs is displayed instead of the DOT BP
mask. The area used to extract the NBP sample is also indicated for the first repetition.  A visual inspection shows that almost all
features identified as G-band bright points (red contours) are
cospatial with polarization signals above the noise. A single
polarization patch usually contains a few BPs, and the majority of
polarization patches show BPs at least in one of the repetitions. Most magnetic
signals remain visible during the full observation run (1 hour).  This
 agrees with the interpretation that brightenings in the G
band are due to the presence of magnetic fields, while the high
temporal variability of the BPs is due to changes in the magnetic
field configuration.

The sunspot canopy (visible near the upper edge of the FOV) clearly
extends beyond the white-light boundary of the spot. The canopy is not
an area of enhanced G-band intensity because its magnetic fields are
very inclined and do not lead to a sufficient downward shift of
the optical depth scale.

\begin{figure*}
\caption{Same as Fig.~\ref{alimapss1}, for the last four repetitions
of the scan. \label{alimapss2}}
\end{figure*}
\end{appendix}
\end{document}